\RequirePackage{lineno} 

\documentclass[twocolumn,preprintnumbers,amsmath,amssymb,nofootinbib]{revtex4}

\usepackage{graphicx}
\usepackage{dcolumn}
\usepackage{bm}
 
\usepackage{epstopdf}

\def\bea{\begin{eqnarray}}
\def\eea{\end{eqnarray}}

\def\pp{\mbox{$p$-$p$}}

\def\pa{\mbox{$p$-A}}

\def\ppb{\mbox{$p$-Pb}}
\def\pw{\mbox{$p$-W}}
\def\pbe{\mbox{$p$-Be}}
\def\pti{\mbox{$p$-Ti}}
\def\ph{\mbox{$p$-H}}

\def\pn{\mbox{$p$-N}}
\def\aa{\mbox{A-A}}
\def\nn{\mbox{N-N}}

\def\pt{$p_t$}
\def\mt{$m_t$}
\def\yt{$y_t$}
\def\yz{$y_z$}

\def\nch{$n_{ch}$}
\def\mmpt{$\bar p_t$}

\begin{document} 

\setlength{\pdfpagewidth}{8.5in}
\setlength{\pdfpageheight}{11in}

\setpagewiselinenumbers
\modulolinenumbers[5]

\addtolength{\footnotesep}{-10mm}\

\preprint{version 3.0\textsl{}}                                             

\title{Analysis of identified-hadron spectra from fixed-target $\bf p$-A collisions\\ and the nature of the Cronin effect
}

\author{Thomas A.\ Trainor}\affiliation{University of Washington, Seattle, WA 98195}


\date{\today} 

\begin{abstract}

In this study fixed-target spectra obtained by the Chicago-Princeton (C-P) collaboration at Fermilab in the mid seventies are analyzed with a two-component spectrum model (TCM) that has been applied successfully to a number of collision systems at the RHIC and LHC in the past. It is from C-P data that the Cronin effect was first inferred. TCM analysis leads to factorization of collision-energy and target A dependences. Over the energy range of C-P data energy dependence is restricted to model-function shapes on $p_t$ whereas A dependence is restricted to particle densities for three hadron species and their antiparticles. A dependence for soft and hard components varying {\em separately} as power laws $A^{\alpha_s}$ and $A^{\alpha_h}$ with {\em fixed} exponents is a central finding of this study. The trends $A^{\alpha(p_t)}$ inferred by the C-P collaboration resulted from treating spectra as monolithic which confuses the distinct $p_t$ (model functions) and A (particle densities) dependences. The Cronin effect resulting from that confusion is easily explained in a TCM context. Power-law trends for pions at 25 GeV are quantitatively compatible with trends at LHC energies. The relation of exponents $a_s$ and $a_h$ to $p$-A centrality is examined in detail. A Glauber model of $p$-A centrality seems invalid.

\end{abstract}

\maketitle

\section{Introduction}

Various data presentation styles and interpretations associated with output from the Relativistic Heavy Ion Collider (RHIC) and Large Hadron Collider (LHC) have been employed in recent years to claim formation of a quark-gluon plasma or QGP in small collision systems -- \pp\ and \pa\ collisions~\cite{gardim,alicestrange,phenix,cmsridge,alippss}. In some cases such conclusions arise from misinterpreting phenomena associated with jet production as manifestations of flow: e.g.\ ``hardening'' of \pt\ spectra with increasing event multiplicity or \nch\ and interpretation of certain Fourier coefficients as representing elliptic and other azimuth-varying flows~\cite{ppquad}.

Regarding jet production and search for possible modification thereof by QGP formation, {\em nuclear modification factors} (NMFs) have been commonly employed: A spectrum ratio is formed from some object spectrum (e.g.\ for a more-central event class) and a reference spectrum (e.g.\ from \pp\ collisions or a peripheral event class). The spectrum ratio is rescaled by an estimator (e.g.\ number of binary \nn\ collisions $N_{bin}$ or target thickness $A^{1/3}$) such that for the case of no jet modification the  NMF is expected to be consistent with unity at higher \pt. For \aa\ collisions reduction below unity at higher \pt\ is interpreted to indicate jet ``quenching'' as confirming QGP formation. For smaller collision systems increase above unity at intermediate \pt\ with subsequent decrease (termed the ``Cronin effect'') complicates interpretation of NMFs.

In previous studies identified-hadron spectra from 5 TeV \ppb\ collisions~\cite{ppbpid,pidpart1,pidpart2} and 13 TeV \pp\ collisions~\cite{pppid,ppsss} were analyzed with a two-component (soft+hard) model (TCM) so as to isolate jet (hard) and nonjet (soft) contributions, with data-model agreement at the level of statistical uncertainties. Evolution of jet structure for several hadron species is accurately described with a simple few-parameter model. No evidence for flows or other QGP-associated phenomena has emerged.

In this study the TCM is applied to spectrum data produced by the Chicago-Princeton (C-P) collaboration at the Fermi National Accelerator Lab (FNAL) in the mid seventies at COM energies near 25 GeV~\cite{cronin10,cronin0}. Those data are of particular interest because the collision energies are just above the threshold for jet production as determined by extrapolation from RHIC and LHC energies and because the Cronin effect as such originates from those data, J. Cronin being a C-P collaboration member.

In what follows the C-P data are introduced and converted from differential cross sections to particle densities for clearer interpretation. A PID (identified hadron) TCM for \pa\ collisions is introduced, and its previous application to 5 TeV \ppb\ spectra is briefly described as an introduction to the model. As further introduction the TCM is applied to C-P pion data to obtain collision-energy and target atomic-weight or size A dependence. 

The main analysis then consists of applying those methods to all published C-P data. A major result of the study is the finding that C-P soft and hard particle densities vary with target size A separately as simple power-law dependences $A^{a_{si}}$ and $A^{a_{hi}}$ with {\em fixed} exponents.  Reported \mbox{C-P} exponent trends $\alpha_i(p_t)$ then arise from the interplay between TCM model functions $\hat S_{0i}(p_t)$ and $\hat H_{0i}(p_t)$ and fixed exponents $a_{si}$ and $a_{hi}$. C-P A dependences for pions are quite similar to  pion trends for 5 TeV \ppb\ collisions at the LHC. The relation $a_h \approx 2a_s$ (for pions at least) is explained in terms of TCM analysis of \ppb\ centrality and particle production. 

The Cronin effect associated with NMFs is then explained as resulting from  fixed $A^{a_x}$ trends interacting with  chosen rescaling  factor $N_{bin}$ or $A^{1/3}$ via soft and hard components. Detailed TCM parameter trends show how jet formation evolves with decreasing collision energy just above the threshold for jet production near 10 GeV.

This article is arranged as follows:
Section~\ref{cpintro} introduces the C-P spectrum data.
Section~\ref{spectrumtcm} describes a PID TCM for \pa\ spectra.
Section~\ref{initial} provides an initial analysis of selected C-P data to introduce the overall strategy.
Section~\ref{speceng} presents the energy dependence of \mbox{C-P}  spectrum structure.
Section~\ref{cpadep} presents the target A dependence of \mbox{C-P} particle densities.
Section~\ref{lhctrends}  compares density trends for C-P vs LHC spectra.
Section~\ref{cpratios} discusses spectrum ratios in various combinations and interprets the results in the context of the TCM.
Sections~\ref{disc} and~\ref{summ} present discussion and summary.

\section{C-P spectrum introduction} \label{cpintro}

References~\cite{cronin10,cronin0} report Chicago-Princeton results from fixed-target experiments at the FNAL using proton beams of 200, 300 and 400 GeV (COM energies approximately 19.4, 23.8 and 27.4 GeV) incident on several targets including H$_2$, Be, Ti and W. In the present study published C-P differential cross-section spectra are converted to the form $d^2n_i/p_t dp_t dy_z$ for hadron species $i$. The conversion is essential in order to better interpret C-P spectrum data. There are two main issues related to the Cronin effect: collision-energy $\sqrt{s}$ dependence and target size A (effective nuclear thickness) dependence.

\subsection{Invariant cross sections}

Published C-P spectra are presented as invariant cross sections in units  cm$^2$/GeV and include as a factor a proton-target cross section $\sigma_{pX}$ -- either the {inelastic cross section} $\sigma_{p\text{N}}$ for \pn\ collisions within nuclei or the {absorption cross section}  $\sigma_{abs} \rightarrow \sigma_{p\text{A}}$ for \pa\ collisions taken as a whole:
\bea \label{oldcross}
E\frac{d^3\sigma_{p\text{A}i}}{dp^3} &=& \sigma_{p\text{X}}\frac{1}{N_{evt}}\frac{Y_i}{p^2 \Delta \Omega \Delta p/p},
\eea
where $Y_i$ is the particle yield for species $i$ into the differential momentum acceptance in the denominator for $N_{evt}$ \pa\ collision events.
In the present study published spectra are converted to measured invariant particle densities on transverse momentum $p_t$ and {\em longitudinal} rapidity $y_z$, distinguished from {\em transverse} rapidity  $y_t$ that is essential for describing jet-related spectrum hard components:
\bea
E\frac{d^3\sigma_{p\text{A}i}}{dp^3} 
&\rightarrow& \bar \rho_{0i} =  \frac{d^2n_i}{p_t dp_t dy_z}.
\eea
The conversion factor is $2\pi \times 10^{27}/\sigma_{pX}$ using whatever cross section $\sigma_{pX}$ in mb {\em was applied to published data}. 
In Refs.~\cite{croninabs,cronin10,cronin0} distinctions are made between per-effective-nucleon cross sections and per-nucleus cross sections via prefactors $\sigma_{p\text{N}}$ vs $\sigma_{p\text{A}}$ respectively in Eq.~(\ref{oldcross}). 

\subsection{Spectrum data}

Reference~\cite{cronin10} (PRD11) reports \pw\ \pt\ spectra for three collision energies. In the data tables spectra are presented as invariant cross sections {\em per-nucleus} ($\sigma_{pX} \rightarrow \sigma_{pA}$) denoted by $Ed\sigma/d^3p$. In figures spectra are instead presented as {\em per-nucleon}  ($\sigma_{pX} \rightarrow \sigma_{p\text{N}}$). Pion, charged-kaon and proton data are acquired concurrently based on Cherenkov detectors. While the pion data are presented as absolute cross sections the kaon and proton data are presented as ratios to pion data, a strategy responding to 50\% uncertainty in the absolute proton beam intensity. 

Establishing a relation between \pa\ cross sections and  corresponding \pp\ cross sections is acknowledged to be model dependent. Several strategies are suggested:
(a) Divide per-nucleus differential cross sections by atomic weight A $\approx 184$ for W. It is demonstrated below that the result is equivalent to rescaling particle densities by $A^{1/3} \rightarrow 5.7$.
(b) Divide per-nucleus cross sections by a  number of {\em effective nucleons} $\sigma_{abs}/\sigma_{pp}$ = 1635/40 = 40.9. That ratio then replaces {absorption cross section} $\sigma_{p\text{A}}$ with fixed inelastic cross section $\sigma_{p\text{N}} \rightarrow \sigma_{pp}$ in Eq.~(\ref{oldcross}) leaving particle densities with  their unique A dependence.
(c) Alternatively  ``The best method, of course, is the one which yields cross sections independent of A [see argument in Ref.~\cite{croninabs}].'' Given results of the present study that expectation appears to be unrealistic.
PRD11 data are taken from its Table I (pions, Figs.~2, 3) Table II (kaons/pions, Fig.~9) and Table III (protons/pions, Fig.~11). Particle/antiparticle ratios appear  in Fig.~13.

Reference~\cite{cronin0} (PRD19) reports measurements of invariant cross section $Ed^3\sigma/d^3p$ on targets H$_2$, Be, Ti and W at fixed-target beam energies 200, 300 and 400 GeV for detected pions, charged kaons and protons. Methods are as reported in PRD11. As above, absolute cross sections are reported for pions whereas kaon and proton data are presented as ratios to pion data. Species ratios are reported because ``...this involved only the information from the Cherenkov counters'' and ratios are thus free from normalization uncertainties. For \ph\ collisions absolute cross sections are also reported for charged kaons and protons consistent with the hadron/pion ratios.

The stated motivation for measurement over a range of target A is ``...to extrapolate to A = 1'' for comparison with \pp\ data. It was expected that ``for the `hard' collisions {only a single nucleon in the nucleus would be involved}. It was therefore a surprise when we found that [assuming a trend $A^{\alpha(p_t)}$]...for all particle types [$\alpha$] grows to be greater than 1.0 at large $p_\perp$.'' Details of spectrum evolution with target A are discussed in Secs.~\ref{cpadep},~\ref{lhctrends} and \ref{cpratios} below.
PRD19 data are taken from its Table I (\ph\ pions, Fig.~2), Table IV (\ph\ protons/pions, Fig.~6), Table V (\ph\ kaons/pions, Fig.~6), Table VII (\ph\ kaons), Table VIII (\ph\ protons),  Table IX (Be, Ti, W pions), Table XI (Be, Ti, W protons/pions, Fig.~14) and Table XII (Be, Ti, W kaons/pions).

\section{PID TCM for $\bf p$-A spectra}  \label{spectrumtcm}

In this section a PID TCM for  5 TeV \ppb\ collisions is introduced, referring to analyses of PID spectra from \ppb\ collisions for lower-mass hadrons~\cite{ppbpid,pidpart1,pidpart2}.  That provides a basis for the C-P TCM described below.  The following is a brief summary of analysis reported in Ref.~\cite{ppbnmf}.

\subsection{PID spectrum TCM definition}   \label{pidspec}

Given a \pt\ spectrum TCM for unidentified-hadron spectra~\cite{ppprd,newpptcm} a corresponding TCM for identified hadrons can be generated by assuming that each hadron species $i$ comprises certain {\em fractions} of TCM soft and hard total particle densities $\bar \rho_{s}$ and $\bar \rho_{h}$ denoted by $z_{si}(n_s)$ and $z_{hi}(n_s)$ so that $\bar \rho_{si} = z_{si}(n_s) \bar \rho_{s}$ and $\bar \rho_{hi} =z_{hi}(n_s) \bar \rho_{h}$. Total densities can in turn be expressed in terms of individual \nn\ densities and nucleon participant and binary-collision numbers as $\bar \rho_s = (N_{part}/2) \bar \rho_{sNN}$ and $\bar \rho_h =N_{bin} \bar \rho_{hNN}$.
The PID spectrum TCM is then 
\bea \label{pidspectcm}
\bar \rho_{0i}(p_t,n_s) &=& S_i(p_t) + H_i(p_t,n_s)
\\ \nonumber &\approx&  \bar \rho_{si} \hat S_{0i}(p_t) +   \bar \rho_{hi} \hat H_{0i}(p_t,n_s),
\eea
where for \pp\ (or \nn) collisions $\bar \rho_h \approx \alpha(\sqrt{s}) \bar \rho_s^2$~\cite{newpptcm} and $\bar \rho_0 = \bar \rho_{s} + \bar \rho_h$ is the measured event-class charge density. $\bar \rho_s \equiv n_s / \Delta \eta$ is then obtained from  $\bar \rho_0$ as the root of $\bar \rho_0 = \bar \rho_{s} + \alpha \bar \rho_s^2$, and $n_s$ serves as an event-class index. \ppb\  spectra are plotted here as densities on \pt\ (as published) vs pion transverse rapidity $y_{t\pi} = \ln[(p_t + m_{t\pi})/m_\pi]$. Products $\bar \rho_{xi}(n_s) = z_{xi}(n_s)\bar \rho_{x}$ in Eq.~(\ref{pidspectcm}) transition to $\bar \rho_{xi}(A)$ to describe the A dependence of C-P spectra.

Unit-normal model functions $\hat S_{0i}(m_t)$ (soft) and $\hat H_{0i}(y_t)$ (hard) are defined as {\em densities on those  argument variables} where they have simple functional forms and then transformed to \pt\ via appropriate Jacobians as necessary.  The carets indicate unit-normal functions.

A key issue for analysis of \pa\ spectra is determination of soft-component model $\hat S_{0i}(m_t,\sqrt{s})$, specifically its collision-energy dependence. The model is defined by
\bea \label{s000}
\hat S_{0i}(m_t) &=& \frac{A_i}{[1 + (m_{ti} - m_i)/n_iT_i]^{n_i}}.
\eea
Constant $A_i$ is determined by the unit-normal condition. The functional form of $\hat S_{0i}(m_t)$ could describe an incompletely equilibrated thermal particle source where exponent $n$ is a measure of heterogeneity or resulting \pt\ variance~\cite{wilk}. However, response to Gribov diffusion is more likely~\cite{gribov,gribov3,gribov2}. Parameter $T_i$ is a fixed value for each hadron species. Criteria for determining soft-component $n_i(\sqrt{s})$ values are described below. 

If plotted in a semilog format as a density on $m_{ti}$ vs pion rapidity \yt\ $\hat S_{0i}(m_t)$ initially follows a $m_i [\cosh(y_{ti})-1]/T$ + constant trend, where $m_{ti}$ and $y_{ti}$ are the proper transverse mass and rapidity for hadron species $i$; see the dashed curve in Fig.~\ref{protonspec} (left). As a result, the obvious corner marking the transition from exponential to power law moves to higher pion \yt\ with increasing hadron mass.

 The unit-integral hard-component model is simply defined on pion rapidity $y_{t\pi} \rightarrow y_t$ as a Gaussian with exponential (on $y_t$) or power-law (on $p_t$) tail at higher \yt\
\bea \label{h00}
\hat H_{0}(y_t) &\approx & B \exp\left\{ - \frac{(y_t - \bar y_t)^2}{2 \sigma^2_{y_t}}\right\}~~~\text{near mode $\bar y_t$}
\\ \nonumber
&\propto &  \exp(- q y_t)~~~\text{for higher $y_t$ -- the tail},
\eea
where the transition from Gaussian to exponential is determined by slope matching (see the macro in Ref.~\cite{hardspec}). Coefficient $B$ is determined by the unit-integral condition. This hard-component model, defined as a density on $y_{t\pi}$, may be transformed to a density on \pt\ or \mt\  via  Jacobian $y_{t\pi} / m_{t\pi} p_t $.  The exponential tail on \yt\ then varies on \pt\ approximately as power law $1/p_t^{q + 2}$ reflecting an underlying jet energy spectrum~\cite{fragevo}. Details are provided in Refs.~\cite{ppbpid,pidpart1,pidpart2}.
For convenience below note that pion \yt\ = 2 corresponds to $p_t \approx 0.5$ GeV/c, \yt\ = 2.7 to 1 GeV/c, \yt\ = 4 to 3.8 GeV/c and \yt\ = 5 to 10 GeV/c.

\subsection{$\bf \hat S_{0i}$ and $\bf \hat H_{0i}$ model-function parameters}

Table~\ref{pidparams} shows 5 TeV \ppb\ TCM model parameters for hard component $\hat H_{0i}(y_t)$ (first three) and soft component $\hat S_{0i}(p_t)$ (last two) reported in Ref.~\cite{pidpart1}. Hard-component parameters vary significantly with hadron species and centrality: Gaussian widths $\sigma_{y_t}$ are greater for mesons than for baryons while exponents $q$ are substantially greater for baryons than for mesons. Peak modes $\bar y_t$ for baryons shift to higher \yt\ with increasing centrality while widths above the mode for mesons decrease. Soft-component model parameter $T$ is independent of collision energy but increases substantially with hadron mass. $\hat S_{0i}(p_t)$ exponent $n$ and hard-component exponent $q$ have substantial collision-energy dependence~\cite{alicetomspec}. 

\begin{table}[h]
	\caption{TCM model parameters for identified hadrons from 5 TeV \ppb\ collisions from Ref~\cite{pidpart1}: hard-component parameters $(\bar y_t,\sigma_{y_t},q)$ and soft-component parameters $(T,n)$. These values apply to the  mid-central \ppb\ event class. 
	}	\label{pidparams}
	\begin{center}
		\begin{tabular}{|c|c|c|c|c|c|} \hline
			& $\bar y_t$ & $\sigma_{y_t}$ & $q$ & $T$ (MeV) &  $n$  \\ \hline
			$ \pi^\pm $     &  $2.43\pm0.01$ & $0.62\pm0.01$ & $4.2\pm0.2$ & $145\pm3$ & $8.5\pm0.5$ \\ \hline
			$K^\pm$    & $2.65\pm0.03$  & $0.57\pm0.01$ & $4.1\pm0.2$ & $200\pm10$ & $14\pm3$ \\ \hline
			$p,\,\bar p$        & $2.97\pm0.03$  & $0.47\pm0.01$ & $4.9\pm0.2$ & $200\pm10$ & $14\pm3$ \\ \hline
		\end{tabular}
	\end{center}
\end{table}

The entries for \ppb\ collisions in Table~\ref{pidparams} define a {\em fixed} TCM reference independent of centrality that describes the {\em mid-central} event class (wherein $\bar y_t \approx 2.95$ for baryons)~\cite{pidpart1}. In Refs.~\cite{pidpart2,pppid} variation of some hard-component model parameters is determined so as to describe all event classes within statistical uncertainties (e.g.\ see Fig.~4 of Ref.~\cite{pidpart2}). Required variations are linear on hard/soft ratio $x(n_s) \nu(n_s)$ (where $x$ is the \nn\ hard/soft ratio $\bar \rho_{hNN}/\rho_{sNN}$ and  $\nu$ is the number of binary collisions per participant nucleon pair): with increasing $x\nu$ hard-component modes $\bar y_t(n_s)$ shift to higher \yt\ for baryons while (for \ppb\ but not \pp) hard-component widths above the mode decrease for mesons. 

\begin{figure}[h]
	\includegraphics[width=3.3in]{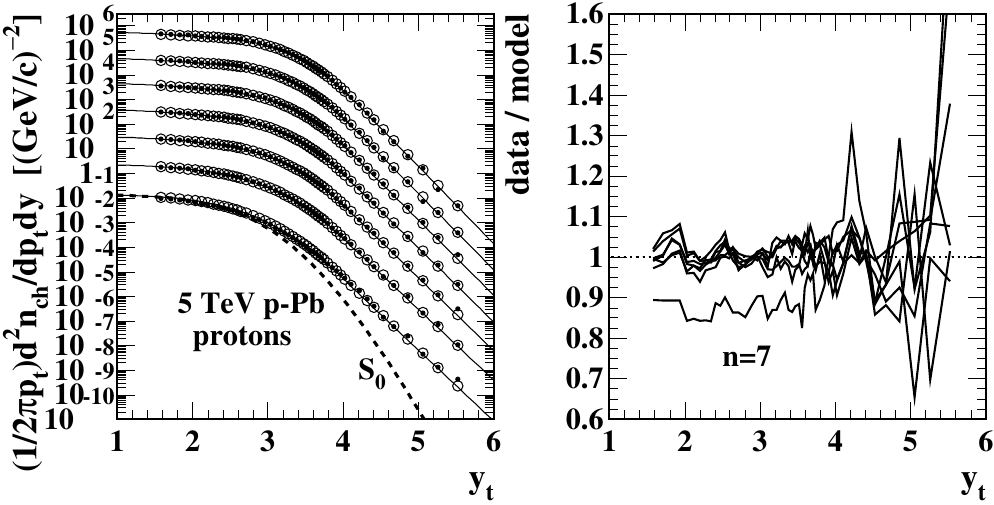}
	\caption{\label{protonspec}
		Left: Proton spectra (solid dots) for seven event classes from 5 TeV \ppb\ collisions reported in Ref.~\cite{alicenucmod}. The TCM is open circles and curves.
		Right: Corresponding data/model ratios. The data spectra are described within statistical uncertainties except for event class $n = 7$.	
	} 
\end{figure}

Figure~\ref{protonspec} (left) shows proton spectrum data (dots) from Ref.~\cite{alicenucmod} compared to the variable TCM (circles and curves) from Ref.~\cite{ppbnmf}. The right panel shows data/model ratios. The most-peripheral event class for $n=7$ is singled out because of the substantial bias common to low-\nch\ data from \pp\ and \ppb\ collisions~\cite{ppbpid,alicetomspec}.

\begin{figure}[h]
	\includegraphics[width=3.3in]{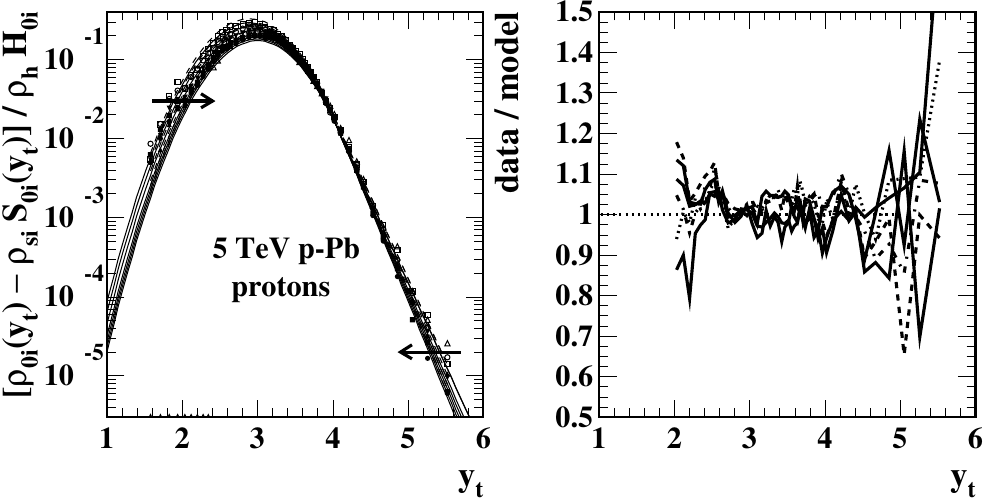}
	\caption{\label{protonhc}
		Left: Data hard components (points) corresponding to proton spectra in Fig.~\ref{protonspec} (left). The curves are variable-TCM hard components. Event class $n=7$ is not shown. The proton hard-component mode shifts to higher \yt\ with increasing \ppb\ centrality coordinated with decreasing amplitude $z_{hi}(n_s)$ so the exponential tail remains stationary near \yt\ = 4.2. At higher \yt\ exponent $q$ increases with softening of the exponential tail. Arrows indicate the trend with increasing centrality.
		Right: Data/model ratios demonstrate that the data are described within statistical uncertainties.
	} 
\end{figure}

Figure~\ref{protonhc} (left) shows proton hard-component results (points) compared to the variable TCM (curves). Note that variation of hard-component structure relative to a {\em fixed} TCM represents all the new spectrum information carried by these data and summarizes completely what could be called ``jet modification'' in \ppb\ collisions.  The right panel shows data/model ratios, demonstrating that data are described accurately within statistical uncertainties.
These items from Ref.~\cite{ppbnmf} serve to illustrate typical results for the PID TCM applied to \pa\ data. The same method is applied to C-P PID spectra below.

\section{C-P spectrum Initial analysis} \label{initial}

This section describes initial analysis of C-P spectra as orientation for the main analysis. A TCM for $\pi^+$ spectra from \pw\ collisions at three energies is first presented to demonstrate energy dependence of soft and hard components. Then TCM A dependence of $\pi^+$ spectra for four targets at 400 GeV is described. The text refers forward to specific numerical results in later sections. In particular, TCM factorization of collision-energy $\sqrt{s}$ and target A dependence per Eq.~(\ref{crontcm}) is emphasized.
In the main analysis below all available C-P data are processed.

\subsection{$\bf p$-A spectrum -- collision-energy $\bf \sqrt{s}$ dependence} \label{cronedep}

\begin{figure}[h]
	\includegraphics[width=1.65in]{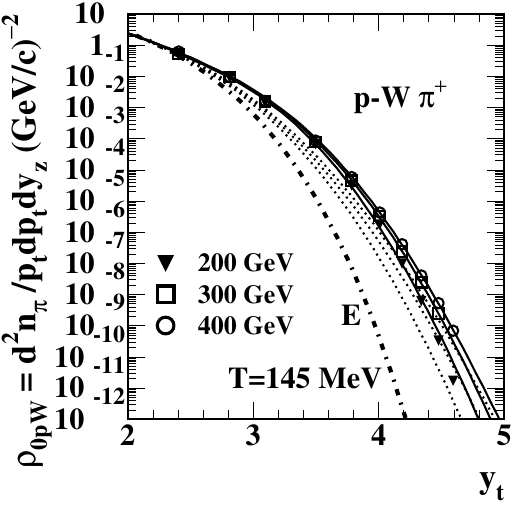}	
\includegraphics[width=1.65in]{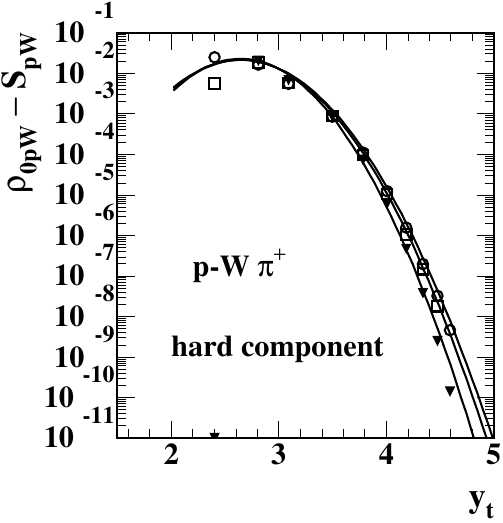}
	\caption{\label{softnx}
		Left:  $\pi^+$ spectra (points) extending to \pt\ $\approx 7$ GeV/c from $p$-W collisions at three fixed-target energies corresponding to $\sqrt{s} = 19.4$, 23.8 and 27.4~\cite{cronin10}.  The dotted and solid curves are the TCM for those collision systems.	
		Right: Data (points) and TCM (curves) spectrum hard components (jet fragments) inferred as described in the text. 
	}  
\end{figure}

Figure~\ref{softnx} (left) shows $\pi^+$ spectra from $p$-W collisions (points) at three fixed-target  beam energies corresponding to $\sqrt{s}\approx  19.4$, 23.8 and 27.4 GeV~\cite{cronin10}. Soft-component models $\hat S_{0i}(p_t,\sqrt{s})$ for $p$-W are maintained identical {\em in shape} to those for \pp\ spectrum data with exponents $n(\sqrt{s})$ in Table~\ref{2aanparamsq}. The dash-dotted curve E is a Boltzmann exponential ($1/n \rightarrow 0$) with $T = 145$ MeV that describes {\em low}-\pt\ pion data for all collision energies significantly above $\sqrt{s} \approx 10$ GeV. Dotted curves are defined by $ S_{p\text{W}}(p_t) = \bar \rho_{s\pi^+} \hat S_{0\pi^+}(p_t)$. Soft charge density $\bar \rho_{s\pi^+}$ is as shown in Table~\ref{pwdensities}. Solid curves through data include hard components inferred from data at right. 

Figure~\ref{softnx} (right) shows spectrum hard components inferred as $H_{p\text{W}}(p_t) = \bar \rho_{0\text{pW}}(p_t) - S_{p\text{W}}(p_t)$ transformed with an appropriate Jacobian to transverse rapidity \yt\ where spectrum hard components have a simple form (approximate Gaussian). As noted, the $S_{p\text{W}}(p_t,\sqrt{s}) = \bar \rho_{si} \hat S_{0i}(p_t,\sqrt{s})$ for three energies (dotted) in the left panel are identical {\em in form} to those for \pp\ collisions. The data hard components in the right panel are represented by TCM models $\bar \rho_{hi} \hat H_{0i}(y_t,\sqrt{s})$. As for $\bar \rho_{s\pi^+}$, collision-energy variation does not significantly change charge density $\bar \rho_{h\pi^+}$ (Table~\ref{pwdensities}) but does change the model shape at higher \pt. The three solid curves for $p$-W correspond to model widths $\sigma_{y_t}(\sqrt{s})$ for increasing collision energy from Table~\ref{pwwidthsxz}. 

Evolution of the soft component in the left panel may be attributed to Gribov diffusion within dissociating projectile nucleons~\cite{gribov,gribov2} whereas evolution of the hard component in the right panel may be attributed to hardening of the underlying parton spectrum leading to jets~\cite{jetspec2,alicetomspec}. There is no significant shift of pion mode $\bar y_t$ with energy, just as observed at higher energies~\cite{alicetomspec}.

Establishment of $n$ values for $\hat S_{0i}(p_t,\sqrt{s})$ proceeds as follows: For each collision energy soft-component model $\bar \rho_{si} \hat S_{0i}(p_t)$  is matched to \pp\ spectrum data (not shown)  at low \pt\ by adjusting $\bar \rho_{si}$. Relative to \pp\ data spectra where the jet contribution is small $n$ is  adjusted for each energy so the dotted curve falls just below the data point at highest \pt. Those values are then {\em fixed independent of target A}. In Fig.~\ref{softnx} (left) those same model functions (dotted) are then used to obtain \pw\ data hard components (points) appearing in the right panel by subtraction. 

Model curves $\bar \rho_{hi}\hat H_{0i}(y_t,\sqrt{s})$ appear as  solid curves at right that accommodate data at and above the mode. Optimized hard-component models $\bar \rho_{hi} \hat H_{0i}(y_t,\sqrt{s})$ are then combined with $\bar \rho_{si}\hat S_{0}(p_t,\sqrt{s})$ to form complete spectrum models appearing as solid curves in Fig.~\ref{softnx} (left). 

\subsection{$\bf p$-A spectrum -- target size A dependence}

\begin{figure}[h]
	\includegraphics[width=3.3in]{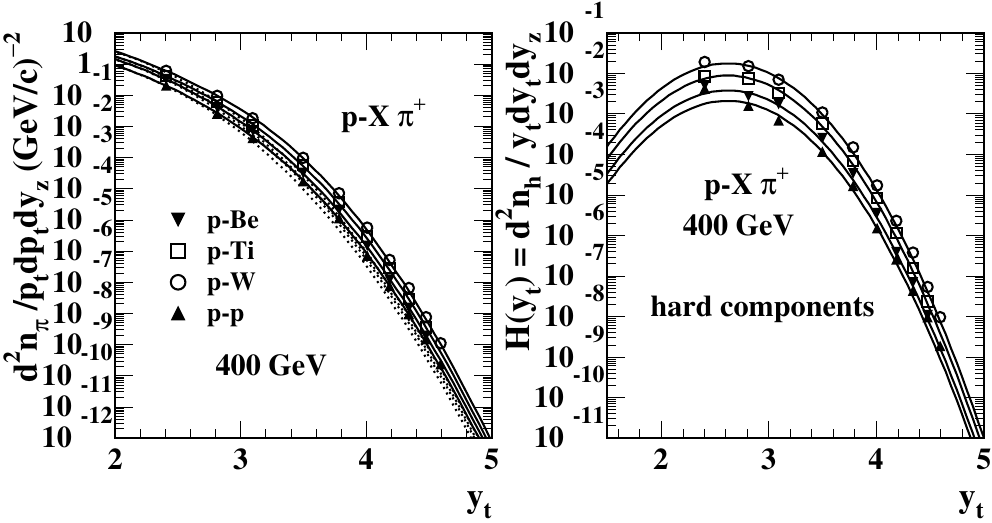}	
		\caption{\label{20d}
		Left: $\pi^+$ spectra from four collision systems (points) with varying target weight A for fixed-target beam energy 400 GeV~\cite{cronin0}. The solid and dotted curves are the corresponding TCM described in the text.
		Right: Data (points) and TCM (curves) hard components inferred as described in the text.
	}  
\end{figure}

For C-P spectra  target size A is relevant rather than \ppb\ centrality index $n_s$ or $\bar \rho_s$. Equation~(\ref{pidspectcm}) can be reformulated in the \pa\ context for hadron species $i$ as
\bea \label{crontcm}
\bar \rho_\text{0pX}(p_t,\sqrt{s},A) &\approx&   S_\text{pX}(p_t,\sqrt{s},A) +   H_\text{pX}(p_t,\sqrt{s},A)~~
\\ \nonumber
&& \hspace{-.6in} \approx \bar \rho_{si}(A) \hat S_{0i}(p_t,\sqrt{s}) + \bar \rho_{hi}(A) \hat H_{0i}(p_t,\sqrt{s}),
\eea
The TCM so defined assumes approximate factorization of target A and energy $\sqrt{s}$ dependences: The {\em shapes} of soft and hard model components are assumed independent of A. As described above, model $\hat S_{0i}(p_t,\sqrt{s})$, i.e.\ parameter $n_i(\sqrt{s})$, is obtained from \pp\ data in Ref.~\cite{cronin0}. Model $\hat H_{0i}(p_t,\sqrt{s})$, i.e.\ parameters $\bar y_{ti}$ and $\sigma_{y_ti}(\sqrt{s})$, is obtained from \pw\ data in Ref.~\cite{cronin10}.  Particle densities $\bar \rho_{xi}(A)$ are then obtained from \pa\ data in Ref.~\cite{cronin0}. The parameter values derived from C-P data are employed in Eq.~(\ref{crontcm}) to generate the solid curves in Fig.~\ref{20d}.

Figure~\ref{20d} (left) shows $\pi^+$ spectra (hadron densities) for four collision systems (points) corresponding to variation of target atomic weight A~\cite{cronin0}. TCM soft components (dotted) use $n$ values determined for 400 GeV \pp\ collisions as described in the previous subsection, with coefficient $\bar \rho_{si}(A)$ values adjusted to accommodate spectrum data at lowest \pt.  400 GeV data provide the best \pt\ coverage.  The resulting TCM describes spectrum data well. 

Figure~\ref{20d} (right) shows hard components (points) inferred as $H_\text{pX}(p_t) = \bar \rho_{0\text{pX}}(p_t) - S_\text{pX}(p_t)$ (refer to Eq.~(\ref{crontcm})) and then transformed to densities on \yt\ with Jacobian $m_t p_t / y_t$. The curves are hard-component model $H_\text{pX}(y_t)$ with coefficient $\bar \rho_{hi}(A)$ values adjusted to best accommodate data.  
Given a TCM analysis method for pion \mbox{C-P} \pt\ spectra the same procedure is applied below  to three hadron species for positive- and negative-signed particles. The energy dependence of TCM model functions is established by comparison of \pp\ and \pw\ spectra for three collision energies. The A dependence of spectra for the highest collision energy is then determined by comparison of 400 GeV \ph, \pbe, \pti\ and \pw\ spectra.

\section{C-P spectrum Energy dependence} \label{speceng}

As noted, the TCM description of C-P energy dependence is based on inferring soft-component properties from \pp\ spectra where the hard component is minimal, and hard-component properties from \pw\ spectra where the hard component is maximal and the soft-component shape has been determined previously from \pp\ data.

\subsection{p-p spectra -- soft components} \label{ppsoft}

\pp\ spectra from Ref.~\cite{cronin0} are plotted here in the left panels as densities on \pt\ (as published)  vs pion \yt\ instead of \pt. The advantages are twofold: At low \yt\ the slopes of the data trends approach zero facilitating estimation of $\bar \rho_{si}$ values. At higher \yt\ a power-law dependence would manifest as a straight line, and hard-component widths above the mode may be  more precisely estimated.

\begin{figure}[h]
	\includegraphics[width=1.65in]{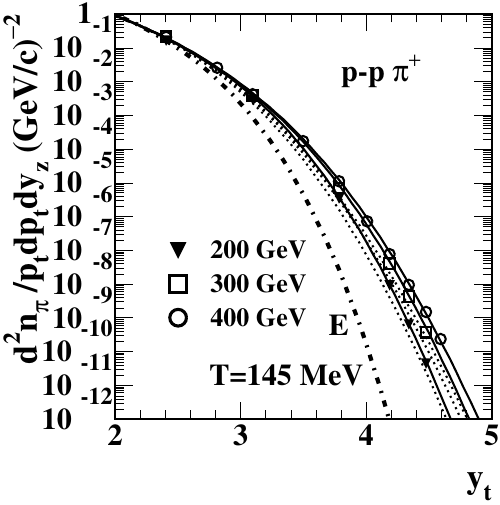}
	\includegraphics[width=1.65in]{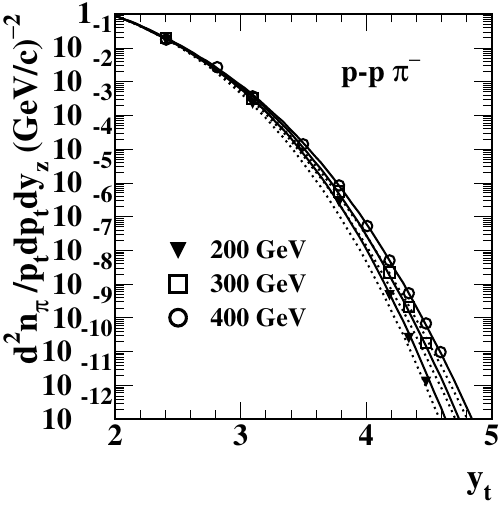}
\put(-144,92) {\bf (a)}
\put(-22,92) {\bf (b)}\\
	\includegraphics[width=1.65in]{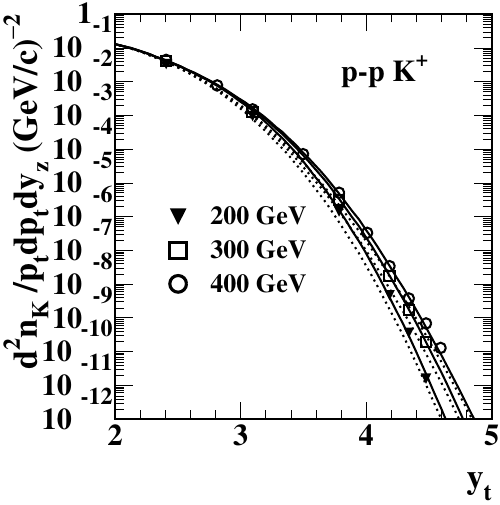}
	\includegraphics[width=1.65in]{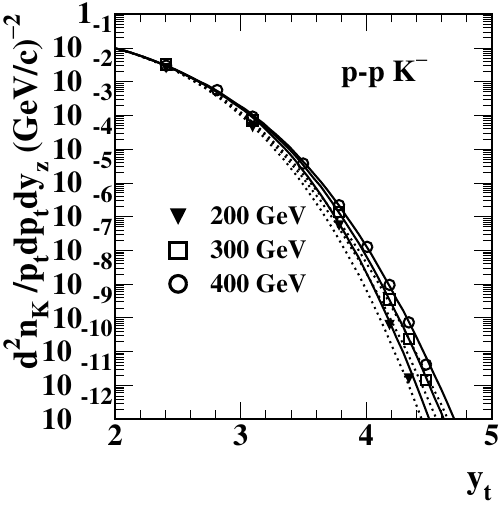}
\put(-144,92) {\bf (c)}
\put(-22,92) {\bf (d)}\\
	\includegraphics[width=1.65in]{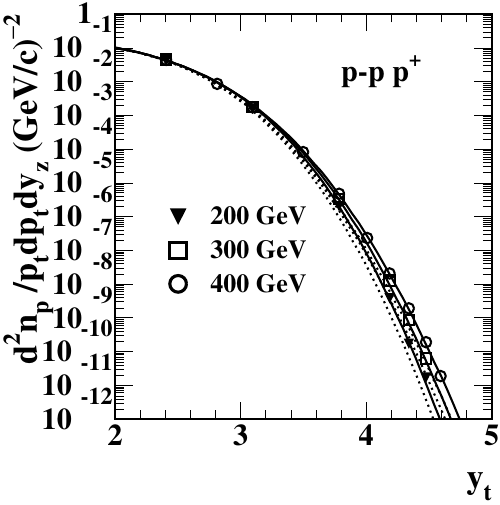}
	\includegraphics[width=1.65in]{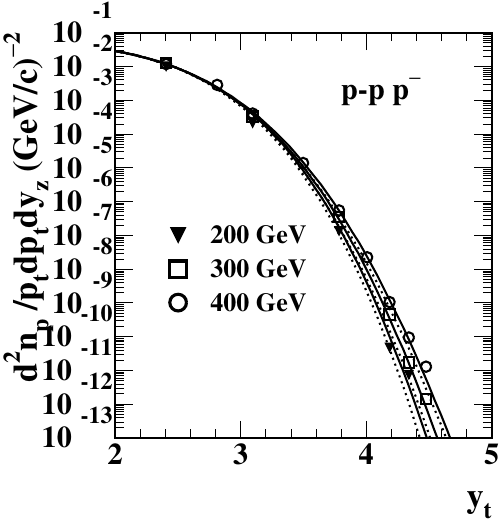}
\put(-144,92) {\bf (e)}
\put(-22,92) {\bf (f)}
	\caption{\label{20aa}
Hadron $h$ spectra (points) from \pp\ collisions for beam energies of 200, 300 and 400 GeV from Ref.~\cite{cronin0} and for $h^+$ (left) and $h^-$ (right). Dotted curves are soft-component models $\bar \rho_{si} \hat S_0(p_t)$ for three collision energies. Solid curves are full TCM including hard components $\bar \rho_{hi} \hat H_0(p_t)$ per Eq.~(\ref{crontcm}).
	} 
\end{figure}

Figure~\ref{20aa} shows C-P positive (left) and negative (right) hadron spectrum data (points) for pions (a,b), charged kaons (c,d) and protons (e,f) from \pp\ collisions for three fixed-target beam energies corresponding to CM energies $\sqrt{s} \approx 19.4$, 23.8 and 27.4 GeV.  The dotted curves are $\hat S_0(p_t,\sqrt{s})$ from Eq.~(\ref{s000}) with charge density $\bar \rho_s$ values in Table~\ref{ppdensities} adjusted to accommodated  lowest-\pt\ data points  and exponent $n$ values in Table~\ref{2aanparamsq} adjusted so that dotted curves pass just below the highest-\pt\ data point for each energy (defining minimum values of $n$ for each spectrum).  Dash-dotted curve E in panel (a) is a Boltzmann exponential ($1/n = 0$) with $T = 145$ MeV (for pions) that describes low-\pt\ data.
The solid curves are the TCM defined by Eq.~(\ref{crontcm}) including soft-component properties inferred here and hard-component properties derived from \pw\ data described below. The combination describes spectrum data within point-to-point uncertainties.

\begin{figure}[h]
	\includegraphics[width=1.6in]{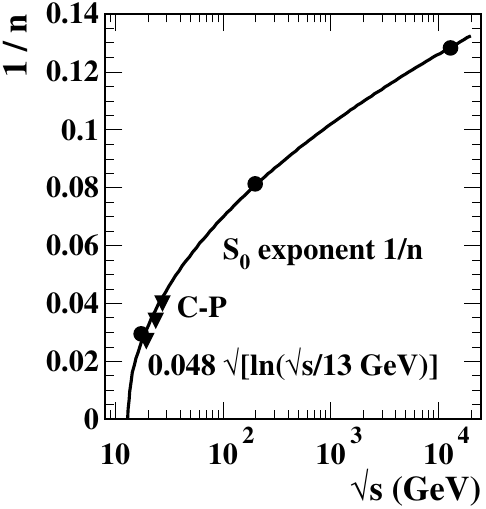}
	\includegraphics[width=1.7in]{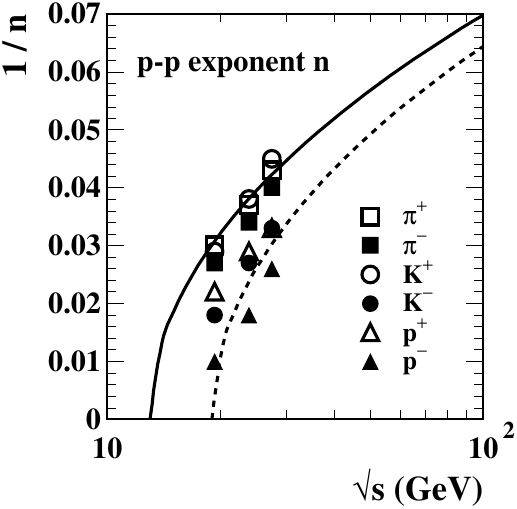}
	\caption{\label{ndep}
		Left: L\'evy exponent $n$ (solid dots) inferred from higher-energy \pp\ collisions as reported in Ref.~\cite{alicetomspec}. Inverted triangles are values inferred from C-P \pp\ $\pi^+$ spectra in Ref.~\cite{ppbnmf}. The solid curve is a proposed trend relating to Gribov diffusion~\cite{alicetomspec,gribov3,gribov2}.
		Right: Values for exponent $n$ from C-P spectra in the present study in an expanded format.
	} 
\end{figure}

Figure~\ref{ndep} (left) shows $\hat S_0(p_t,\sqrt{s})$ exponents $n$ plotted as $1/n$ vs \pp\ collision energy $\sqrt{s}$ that first appeared in Ref.~\cite{alicetomspec}. The plot was later revised in Ref.~\cite{ppbnmf} to include recent results from an initial TCM analysis of C-P PID spectra (inverted triangles) representing $\pi^+$ spectra that shows good agreement with an extrapolated trend (solid curve) derived from higher-energy spectrum data.

Figure~\ref{ndep} (right) shows results from the present study extended to pions, kaons and protons of both signs. Positive kaons are consistent with positive pions. Negative hadrons are lower than positive (softer spectra) and protons fall substantially below kaons. The various trends can be expressed in terms of an effective zero intercept $\sqrt{s_0}$ for curves of the form  $ 1/n\propto \sqrt{\ln(\sqrt{s} / \sqrt{s_0})}$ as reported in Ref.~\cite{alicetomspec}. The effective intercept for C-P pions from the present study is approximately 13 GeV (solid curve). For negative kaons and protons the intercept is approximately 19 GeV (dashed curve). Given an interpretation of Gribov diffusion as the source of nonzero $1/n$ (i.e.\ \pt\ variance) these trends can be interpreted as indicating the effective depth of a parton splitting cascade (diffusion ``time'' interval) preceding formation of hadrons of a specific species: what is the splitting-cascade ``ancestry'' of a particular hadron species?

Table~\ref{ppwidths} shows hard-component widths adjusted to accommodate higher-\pt\ data in Fig.~\ref{20aa}. Soft and hard charge densities are as reported for \pp\ (target H$_2$) collisions in Sec.~\ref{cpadep}. Hard-component modes held fixed independent of energy and target A are as reported in Table~\ref{ppmodes}. 

\begin{table}[h]
	\caption{ \label{ppdensities}
Soft and hard charge densities for three hadron species from C-P \ph\ collisions inferred from spectra in Fig.~\ref{20aa}.
	}
	\begin{center}
		\begin{tabular}{|c|c|c|c|c|} \hline
			X	&  $\bar \rho_{sX^+}$ & $100\bar \rho_{hX^+}$ & $\bar \rho_{sX^-}$  &$100\bar \rho_{hX^-}$ \\ \hline
			$\pi$     &   $0.52\pm0.05$ & $0.60\pm0.06$ & $0.52\pm0.05$ & $0.44\pm0.05$  \\ \hline
			K      &  $0.055\pm0.005$ & $0.25\pm0.03$ & $0.041\pm0.005$ & $0.12\pm0.02$ \\ \hline	
			p     &   $0.042\pm0.005$ & $0.15\pm0.02$ & $0.012\pm0.002$ & $0.012\pm0.002$  \\ \hline
		\end{tabular}
	\end{center}
\end{table}

\begin{table}[h]
	\caption{Soft-component L\'evy exponent $n$ from \pp\ collisions vs  CM energy $\sqrt{s}$. 
	} \label{2aanparamsq}
	\begin{center}
		\begin{tabular}{|c|c|c|c|c|c|c|} \hline
			E (GeV)	& $n_{\pi^+}$ & $n_{\pi^-}$ & $n_{K^+}$ & $n_{K^-}$   & $n_{p^+}$ & $n_{p^-}$   \\ \hline
			19.4    &  $33\pm2$  &  $37\pm2$  & $35\pm2$ & $55\pm5$  & $45\pm5$ & $100\pm10$ \\ \hline
			23.8       & $27\pm2$  & $29\pm2$  & $26\pm2$ & $37\pm5$  & $34\pm3$  & $60\pm5$ \\ \hline	
			27.4     &  $25\pm2$  & $25\pm2$  & $22\pm2$  & $30\pm2$  & $30\pm2$ & $38\pm4$\\ \hline
		\end{tabular}
	\end{center}
\end{table}

\begin{table}[h]
	\caption{Hard-component widths $\sigma_{y_t}$ for \pp\ spectra. Uncertainties are about 3\%.
	} \label{ppwidths}
	\begin{center}
		\begin{tabular}{|c|c|c|c|c|c|c|} \hline
			$\sqrt{s}$ (GeV)	& $\sigma_{\pi^+}$	& $\sigma_{\pi^-}$  & $\sigma_{K^+}$ & $\sigma_{K^-}$   & $\sigma_{p^+}$ & $\sigma_{p^-}$   \\ \hline
			19.4   & 0.325  &  $0.318$  & $0.320$ & $0.290$  & $0.295$ & $0.270$ \\ \hline
			23.8      & 0.350  & $0.345$  & $0.345$ & $0.310$  & $0.320$  & $0.285$ \\ \hline	
			27.4     & 0.365 &  $0.355$  & $0.355$  & $0.325$  & $0.335$ & $0.303$\\ \hline
		\end{tabular}
	\end{center}
\end{table}

\subsection{p-W spectra -- hard components} \label{pwhard}

In the previous subsection $\hat S_0(p_t)$ exponents $n(\sqrt{s})$ for several hadron species are derived from \pp\ spectra reported in Ref.~\cite{cronin0} where the jet contribution is minimal. In this subsection similar models $\hat S_0(p_t,\sqrt{s})$ ($n$ values agree within uncertainties) are used to isolate hard-component energy dependence for \pw\ collisions from Ref.~\cite{cronin10} where jet contributions are greater. Whereas Ref.~\cite{cronin0} includes data for \pw\ as well as other \pa\ systems Ref.~\cite{cronin10} has better \pt\ coverage for \pw\ spectra at lower energies. However, there is an issue with the latter data in that Ref.~\cite{cronin10} spectra are significantly less efficient than Ref.~\cite{cronin0} data at higher \pt. Hard-component widths $\sigma_{y_ti}$ are therefore inferred separately for \pp\ data from Ref.~\cite{cronin0} and \pw\ data from Ref.~\cite{cronin10}. The \pp\ widths are then used for processing \pa\ data from Ref.~\cite{cronin0} to ensure consistent treatment of Ref.~\cite{cronin0} data.

The analysis method proceeds as follows: Soft-component densities $\bar \rho_{si}$ on \pt\ are adjusted for each energy such that soft-component model $S_{pX}(p_t) = \bar \rho_{si}\hat S_{0i}(p_t)$ passes through the lowest data points as in panels (a,c) below. Per Eq.~(\ref{crontcm}) that combination, as defined on data \pt\ values, is subtracted from data spectra to obtain data hard components $H_{pX}(p_t) = \bar \rho_{hi}\hat H_{0i}(p_t)$ (points, transformed to densities on \yt) shown in panels (b,d). Hard-component densities $\bar \rho_{hi}$ are adjusted to accommodate data near the mode. Hard-component modes $\bar y_{ti}$ are held fixed for each hadron species, and widths $\sigma_{y_ti}$ are  adjusted to best describe data at the higher \pt\ values.

\begin{figure}[h]
	\includegraphics[width=1.67in]{alicron20bb}
	\includegraphics[width=1.63in]{alicron20cc}
\put(-145,85) {\bf (a)}
\put(-23,85) {\bf (b)}
\\
	\includegraphics[width=1.67in]{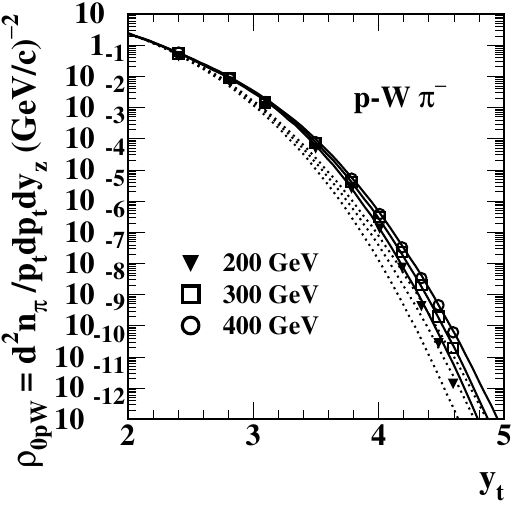}
\includegraphics[width=1.63in]{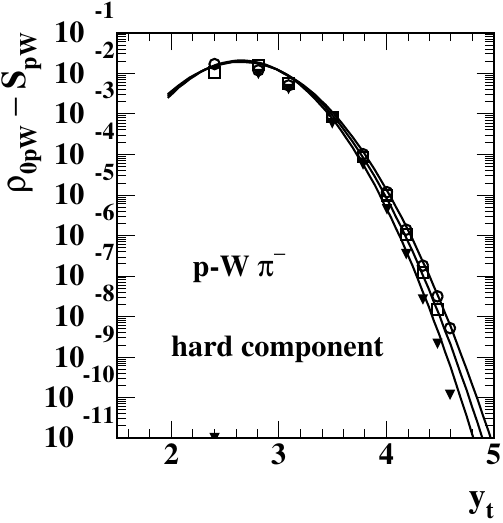}
\put(-145,85) {\bf (c)}
\put(-23,85) {\bf (d)}
	\caption{\label{20bb}
	Left: $\pi^+$ (a) and $\pi^-$ (c) C-P spectra (points) for \pw\ collisions at three energies.  Dotted curves are soft-component models $\bar \rho_{si} \hat S_0(p_t)$ for three collision energies. Dash-dotted curve E is a Boltzmann exponential limit. Solid curves are full TCM including hard components $\bar \rho_{hi} \hat H_0(p_t)$ per Eq.~(\ref{crontcm}).
	Right: Hard components (points) for  $\pi^+$ (b) and $\pi^-$ (d) derived from C-P spectra as described in the text.  The curves are $\bar \rho_{hi} \hat H_{0i}(y_t)$. $\bar \rho_{si}$ and $\bar \rho_{hi}$ values are shown in Table~\ref{pwdensities}.
	} 
\end{figure}

Figure~\ref{20bb} (left) shows $\pi^+$ (a) and $\pi^-$ (c) C-P spectra (points) for \pw\ collisions at three energies. The dotted curves are $\hat S_{0i}(p_t,\sqrt{s})$ obtained from  \pp\ collisions in the previous subsection combined with fixed soft charge density $\bar \rho_{si}$ determined to accommodate the lowest \pt\ points.

Figure~\ref{20bb} (right) shows $\pi^+$ (b) and $\pi^-$ (d) hard components (points) derived by subtracting $\bar \rho_{si}\hat S_{0i}(p_t,\sqrt{s})$ from data in the left panels per Eq.~(\ref{crontcm}). The solid curves through data are TCM hard components $\bar \rho_{hi}\hat H_{0i}(y_t,\sqrt{s})$. 

\begin{figure}[h]
	\includegraphics[width=1.67in]{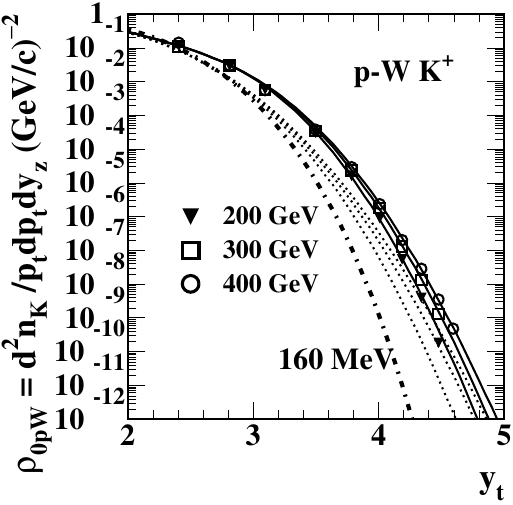}
	\includegraphics[width=1.63in]{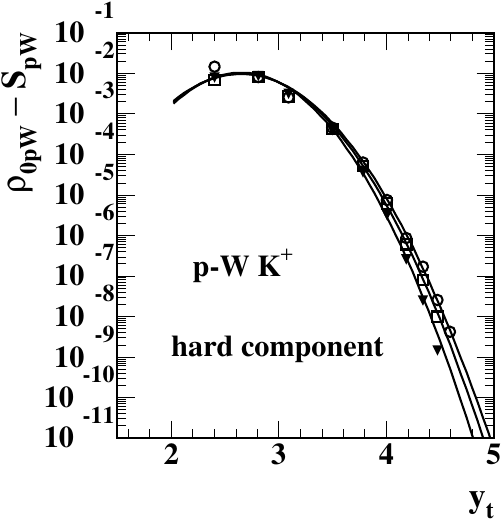}
\put(-145,85) {\bf (a)}
\put(-23,85) {\bf (b)}
\\
	\includegraphics[width=1.67in]{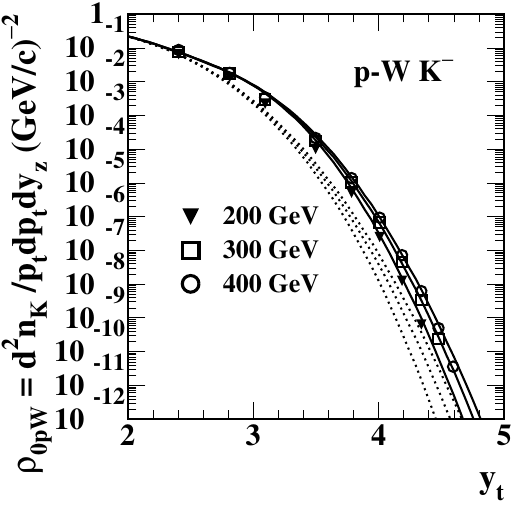}
\includegraphics[width=1.63in]{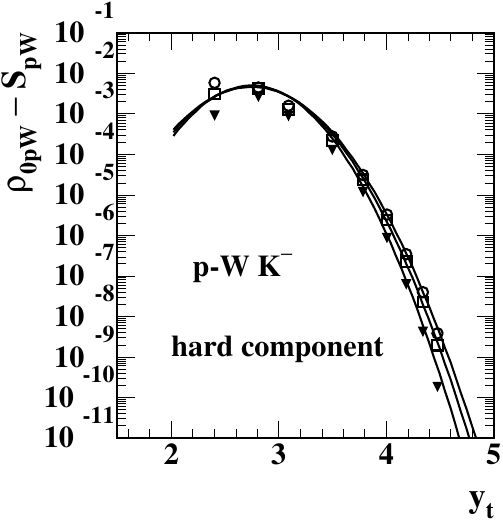}
\put(-145,85) {\bf (c)}
\put(-23,85) {\bf (d)}
	\caption{\label{22bb}
Same as Fig.~\ref{20bb} except for charged kaons.
	}  
\end{figure}

Figure~\ref{22bb} shows comparable results for charged kaons $K^\pm$. Again, the soft-components $\hat S_{0i}(p_t,\sqrt{s})$ are just those derived from \pp\ spectra. While the hard components for $\pi^+$ and $\pi^-$ are quite similar there is a significant difference for charged-kaon hard components in both peak width and charge density $\bar \rho_{hi}$. Refer to Fig.~\ref{xxx} for hard-component widths and Fig.~\ref{trends} for charge densities.

\begin{figure}[h]
	\includegraphics[width=1.67in]{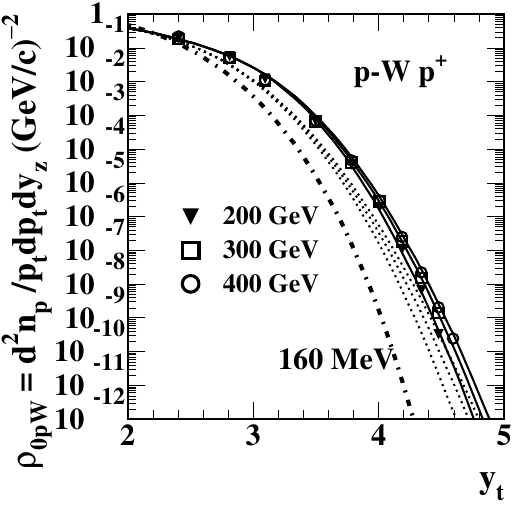}
	\includegraphics[width=1.63in]{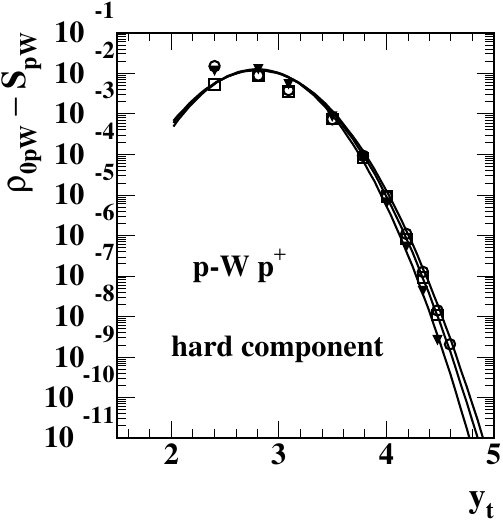}
\put(-145,85) {\bf (a)}
\put(-23,85) {\bf (b)}
\\
	\includegraphics[width=1.67in]{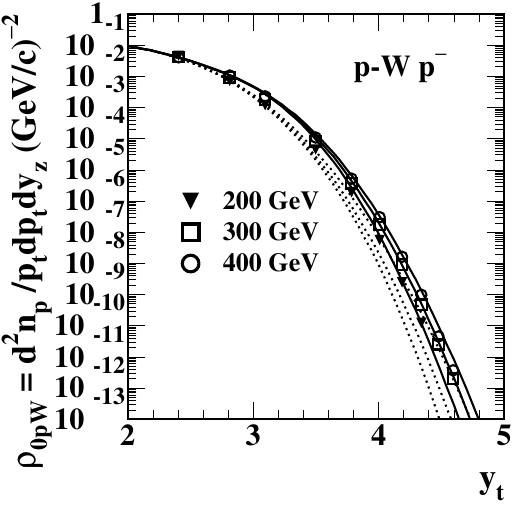}
\includegraphics[width=1.63in]{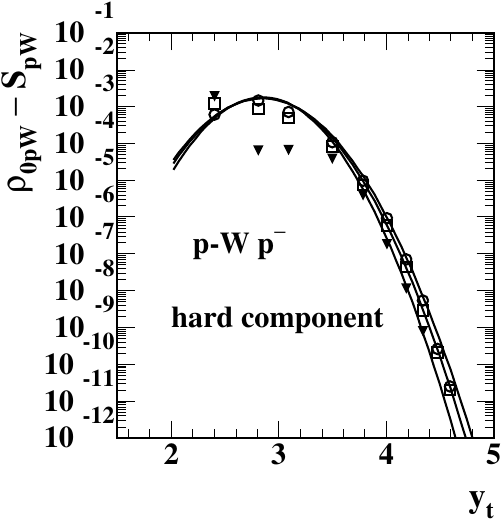}
\put(-145,85) {\bf (c)}
\put(-23,85) {\bf (d)}
	\caption{\label{21bb}
Same as Fig.~\ref{20bb} except for protons and antiprotons.
	}  
\end{figure}

\begin{figure}[h]
\includegraphics[height=1.65in]{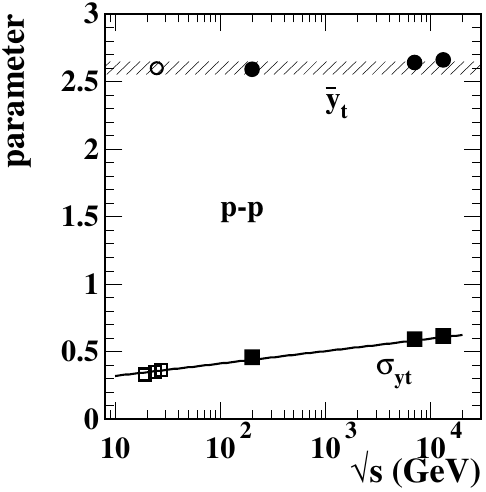}
\includegraphics[height=1.65in]{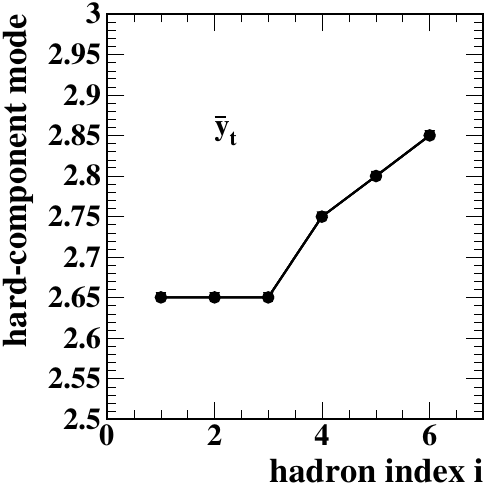}
\put(-144,89) {\bf (a)}
\put(-23,100) {\bf (b)}\\
\includegraphics[height=1.61in]{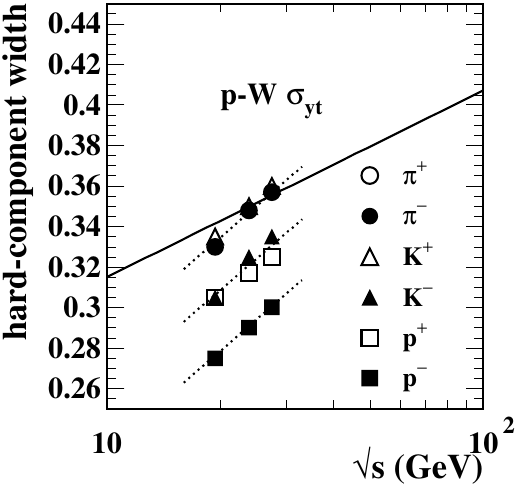}
\includegraphics[height=1.63in]{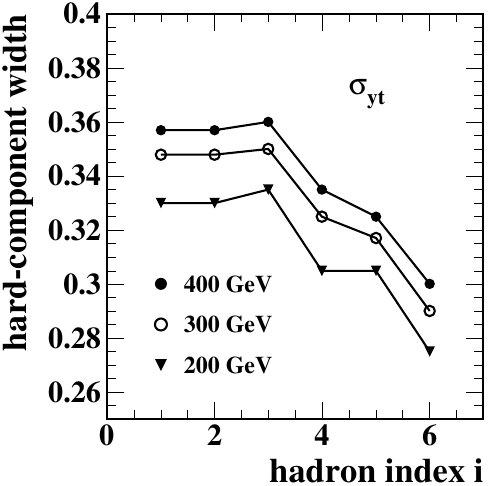}
\put(-148,100) {\bf (c)}
\put(-23,100) {\bf (d)}
	\caption{\label{xxx}
(a) Energy dependence of hard-component parameters $\bar y_{t}$ and $\sigma_{y_t}$ for \pp\ unidentified hadrons  taken from Ref.~\cite{alicetomspec}.
(b) Hard-component modes $\bar y_{ti}$ for Ref.~\cite{cronin0} data vs hadron index $i$ with 1 for $\pi^+$ and 6 for $\bar p$.
(b) Hard-component widths $\sigma_{y_ti}$ vs collision energy for Ref.~\cite{cronin10} \pw\ spectra.
(d) Hard-component widths $\sigma_{y_ti}$ from (c) vs hadron index $i$.
	}  
\end{figure}

Figure~\ref{21bb} shows comparable results for protons $p^\pm \rightarrow p$, $\bar p$. Soft components $\hat S_{0i}(p_t,\sqrt{s})$ are derived from \pp\ spectra as above. For $p^\pm$ there are major differences in both peak widths and charge densities, presumably reflecting specifics of the \pn\ initial state.

Table~\ref{pwdensities} summarizes  \pw\ charge densities. Table~\ref{2aanparams} summarizes soft-component exponents $n$. Table~\ref{pwwidthsxz} summarizes \pw\ hard-component peak widths.  \pw\ hard-component peak modes in Table~\ref{ppmodes} are identical to those for \pp\ spectra. Note that C-P $\hat S_0(p_t)$ exponent $n$ (Fig.~\ref{ndep}) and $\hat H_0(p_t)$ width $\sigma_{y_t}$ (Fig.~\ref{xxx}) vs $\sqrt{s}$ have similar trends vs hadron species independent of A.

\begin{table}[h]
	\caption{ \label{pwdensities}
Soft and hard charge densities for three hadron species from C-P \pw\ collisions inferred from spectra in Figs.~\ref{20bb}, \ref{22bb} and \ref{21bb}.	
}
	\begin{center}
		\begin{tabular}{|c|c|c|c|c|} \hline
			X	&  $\bar \rho_{sX^+}$ & $100\bar \rho_{hX^+}$ & $\bar \rho_{sX^-}$  &$100\bar \rho_{hX^-}$ \\ \hline
			$\pi$     &   $1.40\pm0.10$ & $4.90\pm0.50$ & $1.40\pm0.10$ & $4.30\pm0.40$  \\ \hline
			K      &  $0.120\pm0.02$ & $2.30\pm0.30$ & $0.082\pm0.01$ & $1.05\pm0.10$ \\ \hline	
			p     &   $0.160\pm0.02$ & $2.70\pm0.30$ & $0.039\pm0.05$ & $0.35\pm0.05$  \\ \hline
\end{tabular}
	\end{center}
\end{table}

\begin{table}[h]
	\caption{Soft-component L\'evy exponent $n$ from \pw\ collisions vs $p$-W CM energy $\sqrt{s}$.
	} \label{2aanparams}
	\begin{center}
		\begin{tabular}{|c|c|c|c|c|c|c|} \hline
			$\sqrt{s}$ (GeV)	& $n_{\pi^+}$ & $n_{\pi^-}$ & $n_{K^+}$ & $n_{K^-}$   & $n_{p^+}$ & $n_{p^-}$   \\ \hline
			19.4    &  $33\pm2$  &  $37\pm2$  & $37\pm2$ & $60\pm5$  & $42\pm5$ & $90\pm10$ \\ \hline
			23.8       & $27\pm2$  & $29\pm2$  & $27\pm2$ & $40\pm5$  & $32\pm3$  & $55\pm5$ \\ \hline	
			27.4     &  $23\pm2$  & $25\pm2$  & $23\pm2$  & $32\pm2$  & $27\pm2$ & $36\pm4$\\ \hline
		\end{tabular}
	\end{center}
\end{table}

\begin{table}[h]
	\caption{Hard-component widths $\sigma_{y_t}$ from \pw\ spectra vs $p$-W CM energy $\sqrt{s}$. The widths are plotted in Fig.~\ref{xxx} (c,d) Systematic uncertainties are about 2\%.
	} \label{pwwidthsxz}
	\begin{center}
		\begin{tabular}{|c|c|c|c|c|c|c|} \hline
			$\sqrt{s}$ (GeV)	& $\sigma_{\pi^+}$	& $\sigma_{\pi^-}$  & $\sigma_{K^+}$ & $\sigma_{K^-}$   & $\sigma_{p^+}$ & $\sigma_{p^-}$   \\ \hline
			19.4   & 0.330  &  $0.330$  & $0.335$ & $0.305$  & $0.305$ & $0.275$ \\ \hline
			23.8      & 0.345  & $0.345$  & $0.350$ & $0.320$  & $0.315$  & $0.290$ \\ \hline	
			27.4     & 0.352 &  $0.354$  & $0.360$  & $0.330$  & $0.325$ & $0.295$\\ \hline
		\end{tabular}
	\end{center}
\end{table}

\begin{table}[h]
	\caption{TCM hard-component modes $\bar y_t$ applied consistently to all C-P spectra. The uncertainties are about 2\%.
	} \label{ppmodes}
	\begin{center}
		\begin{tabular}{|c|c|c|c|c|c|} \hline
			 $\bar y_{t\pi^+}$	& $\bar y_{t\pi^-}$  & $\bar y_{tK^+}$ & $\bar y_{tK^-}$   & $\bar y_{tp^+}$ & $\bar y_{tp^-}$   \\ \hline
		 2.65 &  $2.65$  & $2.65$  & $2.75$  & $2.78$ & $2.85$\\ \hline
		\end{tabular}
	\end{center}
\end{table}

\section{C-P spectrum A dependence} \label{cpadep}

This section presents  target A dependence of C-P spectra, specifically  A and species-$i$ dependence of soft $\bar \rho_{si}(A)$ and hard $\bar \rho_{hi}(A)$ particle densities. The inferred trends are simple power laws $A^{a_{xi}}$ used in Secs.~\ref{lhctrends} and~\ref{cpratios} to better understand the structure of spectrum ratios, so-called nuclear modification factors and the Cronin effect. Given lack of significant energy dependence in Fig.~19 of Ref.~\cite{cronin0} the present A-dependence study focuses on 400 GeV data where \pt\ coverage is most complete.

\subsection{p-A analysis procedure}

The TCM description of C-P A dependence is based on the factorization described by Eq.~(\ref{crontcm}): the soft-component model $\hat S_{0i}(p_t)$ shape for a given hadron species $i$ (of specific sign) and collision energy is the same for all target A (confirmed for example for \pp\ vs \pw). Soft-component charge densities $\bar \rho_{si}(A)$ are first adjusted to accommodate lowest-\pt\ data points. Combined spectrum soft-component model $S_{i}(p_t) = \bar \rho_{si}(A) \hat S_{0i}(p_t)$ is then subtracted from data spectra to obtain data hard components $H_i(p_t) =\bar \rho_{hi}(A) \hat H_{0i}(p_t)$. From those results hard-component charge densities $\bar \rho_{hi}(A)$ are obtained by accommodating hard-component $H_i(p_t)$ data near the peak mode. Estimates of $\bar \rho_{si}(A)$ and $\bar \rho_{hi}(A)$ should be insensitive to model details: $\bar \rho_{si}(A)$ is inferred where $\hat S_{0i}(p_t)$ is indistinguishable from a Boltzmann exponential, and $\bar \rho_{hi}(A)$ is inferred where $\hat H_{0i}(p_t)$ is at its maximum.

Model functions vary with species and sign. Soft-component models are consistent with the \pp\ analysis above using 400 GeV  $n$ values summarized in Table~\ref{2aanparamsq}. Hard-component model parameters begin with 400 GeV \pw\ values inferred in Sec.~\ref{pwhard} from Ref.~\cite{cronin10} (PRD11) data. However, due to \pt-dependent differences in particle efficiency that are equivalent to small (few percent) hard-component width differences, here all \mbox{C-P} data are from Ref.~\cite{cronin0} (PRD19) assumed as reference data.  The \pa\ hard-component width data may thus differ significantly from those in  Sec.~\ref{pwhard}. The {\em same} model functions are used for each of four targets A per Eq.~(\ref{crontcm}).

\subsection{400 GeV $\bf p$-A pions}

\begin{figure}[h]
	\includegraphics[width=3.3in]{alicron20d}
\put(-145,105) {\bf (a)}
\put(-23,105) {\bf (b)}\\
	\includegraphics[width=3.3in]{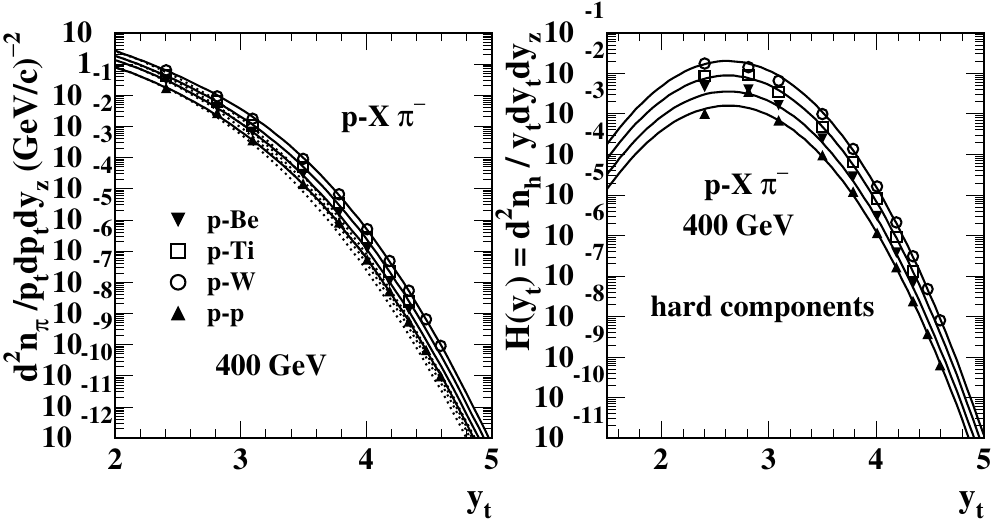}
\put(-145,105) {\bf (c)}
\put(-23,105) {\bf (d)}
	\caption{\label{20dd}
	Left: $\pi^+$ (a) and $\pi^-$ (c) C-P spectra (points) for \pa\ collisions and four targets.  Dotted curves are soft-component $\bar \rho_{si}(A) \hat S_0(p_t)$ all with the same $\hat S_0(p_t)$ for 400 GeV. Solid curves are full TCM including hard components $\bar \rho_{hi}(A) \hat H_0(p_t)$.
	Right: Hard components (points) for  $\pi^+$ (b) and $\pi^-$ (d) derived from C-P spectra as described in the text.  The curves are $\bar \rho_{hi}(A) \hat H_{0i}(p_t)$. $\bar \rho_{si}(A)$ and $\bar \rho_{hi}(A)$ values for pions are shown in Table~\ref{pionaparams}.
	} 
\end{figure}

Figure~\ref{20dd} (left) shows $\pi^+$ (a) and $\pi^-$ (c) C-P spectra (points) for four \pa\ collision systems. The dotted curves are the same 400 GeV soft-component models $\hat S_{0i}(p_t)$ for species $\pi^\pm$ and for 400 GeV beam energy derived in Sec.~\ref{ppsoft} combined with soft-component charge densities $\bar \rho_{si}(A)$ adjusted to accommodate lowest-\pt\ data points.

Figure~\ref{20dd} (right) shows $\pi^+$ (b) and $\pi^-$ (d) hard components (points) derived by subtracting soft-component models $S_i(p_t,A)$ from data in the left panels. The resulting data hard components $H_i(p_t)$ are  then transformed to densities on pion transverse rapidity \yt\ by proper Jacobian. As noted above, hard-component charge densities $\bar \rho_{hi}(A)$ are obtained by accommodating data at the peak mode, and peak widths are adjusted to accommodate data at higher \yt\ (Table~\ref{pwwidthsq}).  The optimized model $H_i(y_t,A) = \bar \rho_{hi}(A) \hat H_{0i}(y_t)$ is shown as the solid curves. Combined soft- and hard-component models (TCM) per Eq.~(\ref{crontcm}) then appear as the solid curves in the left panels.

Table~\ref{pionaparams} summarizes soft and hard charge densities for pions from 400 GeV beam-energy \pa\ collisions.

\begin{table}[h]
	\caption{$p$-A soft-component pion soft-component charge density $\bar \rho_{s\pi^\pm}$ and hard-component charge density $\bar \rho_{h\pi^\pm}$  vs target atomic weight A at 400 GeV.
	} \label{pionaparams}
	\begin{center}
		\begin{tabular}{|c|c|c|c|c|} \hline
			A	&  $\bar \rho_{s\pi^+}$ & $100\bar \rho_{h\pi^+}$ & $\bar \rho_{s\pi^-}$ & $100\bar \rho_{h\pi^-}$   \\ \hline
			H     &   $0.52\pm0.05$ & $0.60\pm0.06$ & $0.52\pm0.05$ &  $0.44\pm0.05$  \\ \hline
			Be       &  $0.82\pm0.08$ & $1.10\pm0.10$ & $0.79\pm0.08$  &$0.85\pm0.08$ \\ \hline	
			Ti     &   $1.02\pm0.10$ & $2.60\pm0.30$ & $1.05\pm0.10$ &  $2.10\pm0.20$ \\ \hline
			W    &  $1.40\pm0.14$  & $5.00\pm0.50$ & $1.40\pm0.14$ & $4.50\pm0.50$ \\ \hline
		\end{tabular}
	\end{center}
\end{table}

\subsection{400 GeV $\bf p$-A charged kaons}

\begin{figure}[h]
	\includegraphics[width=3.3in]{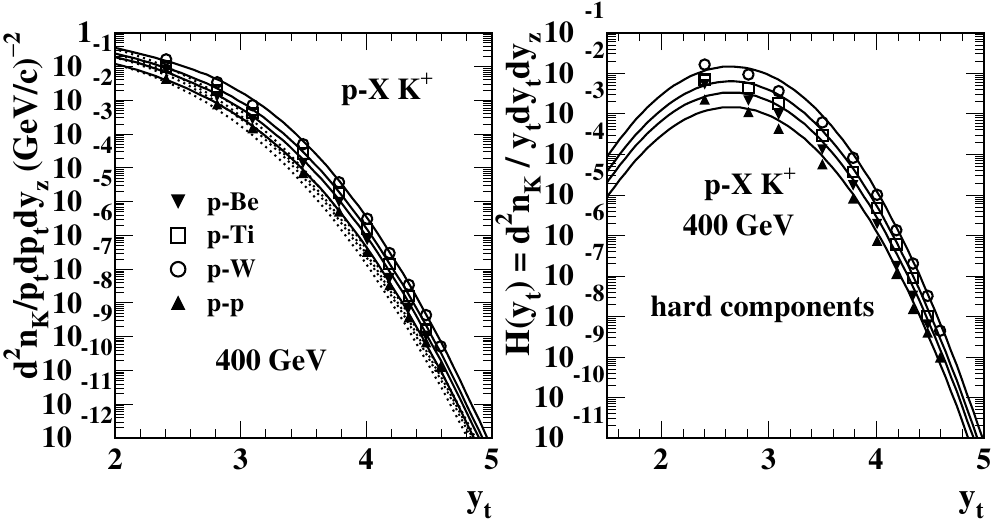}
\put(-145,85) {\bf (a)}
\put(-23,85) {\bf (b)}\\
	\includegraphics[width=3.3in]{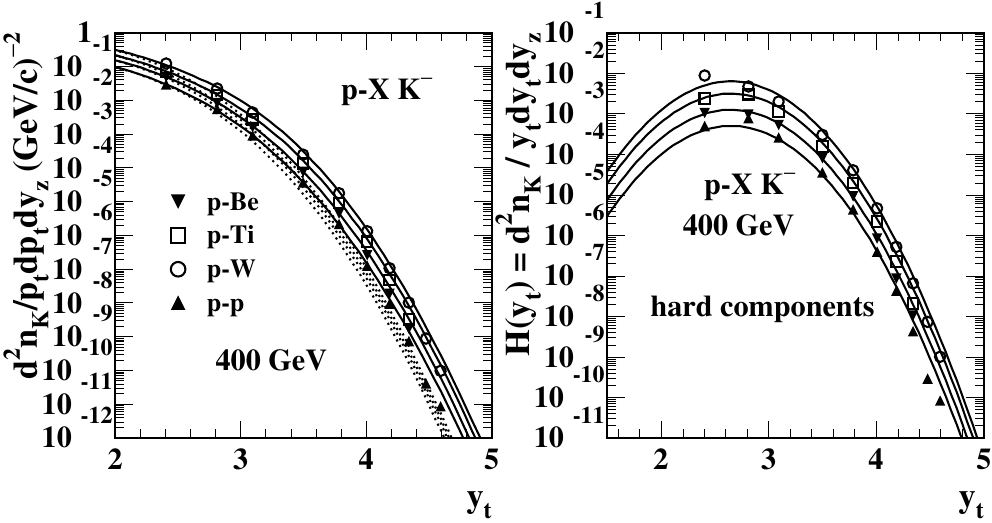}
\put(-145,85) {\bf (c)}
\put(-23,85) {\bf (d)}
	\caption{\label{22dd}
Same as Fig.~\ref{20dd} except for charged kaons.
 $\bar \rho_{si}(A)$ and $\bar \rho_{hi}(A)$ values for kaons are shown in Table~\ref{kaonaparams}.
	} 
\end{figure}

Figure~\ref{22dd} (left) shows $K^+$ (a) and $K^-$ (c) C-P spectra (points) for four \pa\ collision systems. The dotted and solid curves are determined as described above for pions.

Figure~\ref{22dd} (right) shows $K^+$ (b) and $K^-$ (d) hard components (points) derived by subtracting soft-component models $S_i(p_t,A)$ from data in the left panels as described for pions.  The solid curves are determined as described above for pions.

Table~\ref{kaonaparams} summarizes soft and hard charge densities for charged kaons from 400 GeV beam-energy \pa\ collisions.

\begin{table}[h]
	\caption{$p$-A soft-component charge density $\bar \rho_{sK^\pm}$ and hard-component charge density $\bar \rho_{hK^\pm}$ vs target atomic weight A at 400 GeV. 
	} \label{kaonaparams}
	\begin{center}
		\begin{tabular}{|c|c|c|c|c|} \hline
			A	&  $\bar \rho_{sK^+}$ & $100\bar \rho_{hK^+}$ & $\bar \rho_{sK^-}$  &$100\bar \rho_{hK^-}$ \\ \hline
			H     &   $0.055\pm0.06$ & $0.25\pm0.03$ & $0.041\pm0.04$ & $0.12\pm0.01$  \\ \hline
			Be       &  $0.090\pm0.01$ & $0.60\pm0.05$ & $0.065\pm0.06$ & $0.30\pm0.03$ \\ \hline	
			Ti     &   $0.110\pm0.01$ & $1.30\pm0.15$ & $0.100\pm0.01$ & $0.65\pm0.07$  \\ \hline
			W    &  $0.180\pm0.02$  & $2.80\pm0.30$ & $0.140\pm0.02$ &$1.50\pm0.15$  \\ \hline
		\end{tabular}
	\end{center}
\end{table}

\subsection{400 GeV $\bf p$-A protons}

\begin{figure}[h]
	\includegraphics[width=3.3in]{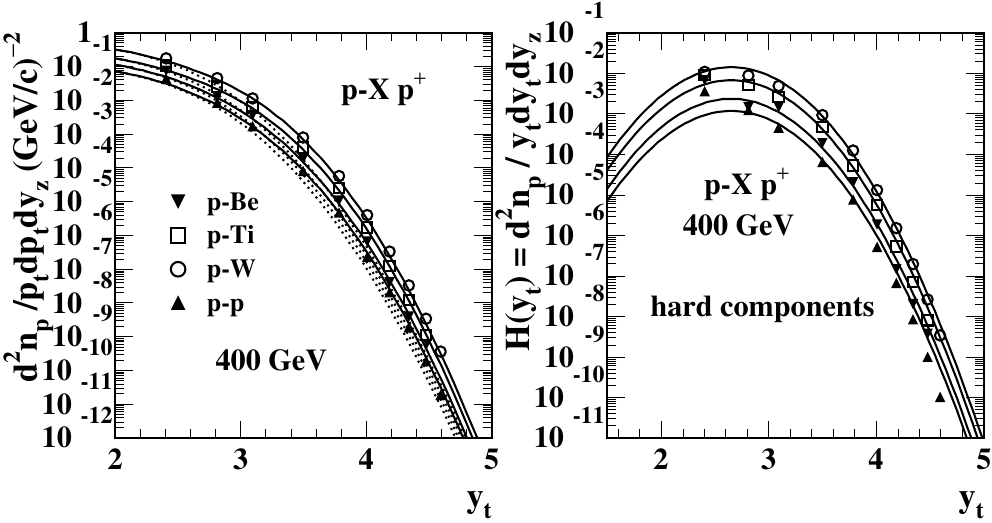}
\put(-145,85) {\bf (a)}
\put(-23,85) {\bf (b)}\\
	\includegraphics[width=3.3in]{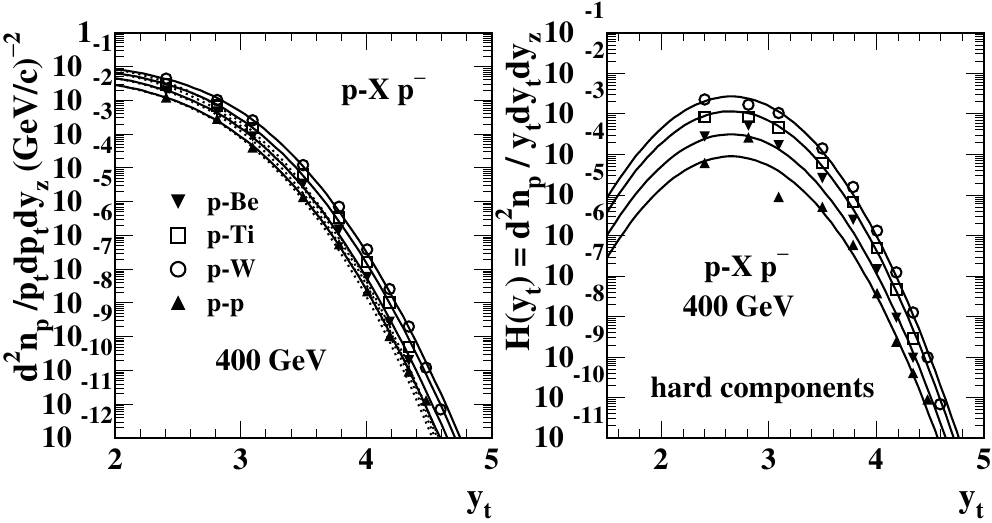}
\put(-145,85) {\bf (c)}
\put(-23,85) {\bf (d)}
	\caption{\label{21dd}
Same as Fig.~\ref{20dd} except for protons and antiprotons.
 $\bar \rho_{si}(A)$ and $\bar \rho_{hi}(A)$ values for protons are shown in Table~\ref{protonaparams}.	
}  
\end{figure}

Figure~\ref{21dd} (left) shows $p^+ \rightarrow p$ (a) and $p^- \rightarrow \bar p$ (c) C-P spectra (points) for four \pa\ collision systems. The dotted and solid curves are determined as described above for pions.

Figure~\ref{21dd} (right) shows $p^+$ (b) and $p^-$ (d) hard components (points) derived by subtracting soft-component models $S_i(p_t,A)$ from data in the left panels as described for pions.  The solid curves are determined as described above for pions.

Table~\ref{protonaparams} summarizes soft and hard charge densities for protons from beam energy 400 GeV \pa\ collisions.

\begin{table}[h]
	\caption{$p$-A soft-component charge density $\bar \rho_{sp^\pm}$ and hard-component charge density $\bar \rho_{hp^\pm}$  vs target atomic weight A at 400 GeV. 
	} \label{protonaparams} 
	\begin{center}
		\begin{tabular}{|c|c|c|c|c|} \hline
			A	&  $\bar \rho_{sp^+}$ & $100\bar \rho_{hp^+}$ & $\bar \rho_{sp^-}$  &$100\bar \rho_{hp^-}$ \\ \hline
			H     &   $0.042\pm0.005$ & $0.15\pm0.02$ & $0.012\pm0.003$ & $0.012\pm0.003$  \\ \hline
			Be       &  $0.055\pm0.006$ & $0.56\pm0.06$ & $0.019\pm0.002$ & $0.05\pm0.005$ \\ \hline	
			Ti     &   $0.084\pm0.01$ & $1.40\pm0.15$ & $0.029\pm0.004$ & $0.17\pm0.02$  \\ \hline
			W    &  $0.136\pm0.015$  & $3.2\pm0.40$ & $0.040\pm0.005$ &$0.40\pm0.05$  \\ \hline
		\end{tabular}
	\end{center}
\end{table}

Soft-component exponents $n$ from \pp\ collisions with 400 GeV beam energy used in the \pa\ study are from from Table~\ref{2aanparamsq}. Hard-component widths $\sigma_{y_t}$ from \pp\ collisions used in the \pa\ study are from Table~\ref{ppwidths}.

\subsection{p-A particle density trends vs A} \label{atrends}

\begin{figure}[t]
	\includegraphics[width=1.676in]{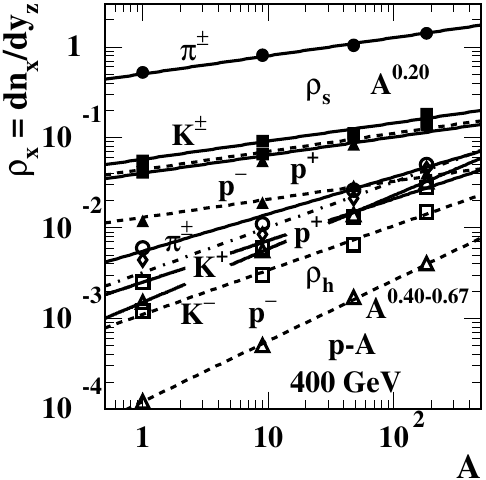}
	\includegraphics[width=1.63in]{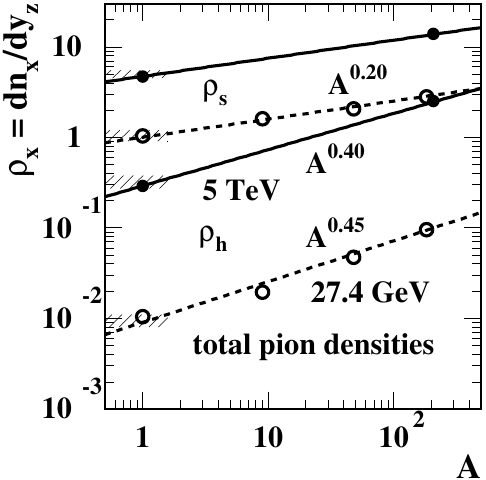}
	\caption{\label{trends}
		Left: $\bar \rho_{si}(A)$ and $\bar \rho_{hi}(A)$ (points) for three hadron species (particles and antiparticles) vs target atomic weight A. The soft-component upper lines for particles (solid) and antiparticles (dashed) vary approximately as $\propto A^{0.20}$.  The hard-component lower lines vary over a range as $\propto A^\text{0.40-0.67}$.
		Right: Trends as in the left panel for C-P total pion densities (open points, dashed lines) and 5 TeV \ppb\ total pion densities (solid dots, solid lines). Given data uncertainties the LHC trends are not significantly different from the C-P trends.
	} 
\end{figure}

Figure~\ref{trends} (left) shows  $\bar \rho_{si}(A)$ and $\bar \rho_{hi}(A)$ data trends (points) for three hadron species and their antiparticles from Tables~\ref{pionaparams}, \ref{kaonaparams} and \ref{protonaparams}. The solid points correspond to  $\bar \rho_{si}(A)$ and open points to $\bar \rho_{hi}(A)$. The solid lines corresponding to particles and dashed lines to antiparticles  reflect power-law trends $A^{a_x}$ with $x = s$ for soft trends and $x=h$ for hard trends. Soft-component trends (upper) are all consistent with $a_s \approx 0.20$ whereas hard-component trends (lower) are consistent with a range of values $a_h \rightarrow \text{0.40-0.67}$ depending on hadron species.

Figure~\ref{trends} (right) shows the trends for C-P {\em total} pion densities (open points, dashed lines) from the present study and the equivalent for 5 TeV \pp\ and \ppb\ data from a TCM analysis with values reported in Ref.~\cite{ppbnmf}, Table I (solid dots, solid lines). The \ppb\ points at A = 208 represent event class $n = 5$ from Ref.~\cite{ppbnmf} that corresponds closely  to NSD \ppb\ as reported by Ref.~\cite{alicenucmod}. The solid points thus represent minimum-bias event ensembles comparable to the C-P data to good approximation.

Immediately notable is the near equivalence of  $A^{a_x}$ trends at 27.4 GeV and 5 TeV for pions. Close correspondence is important because the \nch\ dependence of systems at 5 and 13 TeV has been studied in detail~\cite{ppbpid,pidpart1,pidpart2,newpptcm} and may be used to better understand the C-P data trends as discussed in the next section.

\vskip .2in

Note that Figs.~\ref{ndep} (soft-component model), \ref{xxx} (hard-component model), and \ref{trends} (power-law exponents $a_{xi}$) represent all information from C-P data at the level of statistical uncertainties in an interpretable form. Those parameter values in turn have simple systematic trends, the combination representing {\em lossless data compression}. 

\subsection{C-P analysis compared with a similar study}

Reference~\cite{busza4} reports a comprehensive Fermilab-MIT study of hadron-nucleus collisions with hadron beams at 50, 100 and 200 GeV and targets from hydrogen to uranium. Weak dependence of particle production on target atomic weight A is noted, suggesting that final-state hadrons form well after $h$-A collisions terminate. This Fermilab study is contemporary with the C-P study and it is instructive to compare results from the two.

\begin{figure}[h]
	\includegraphics[width=3.3in]{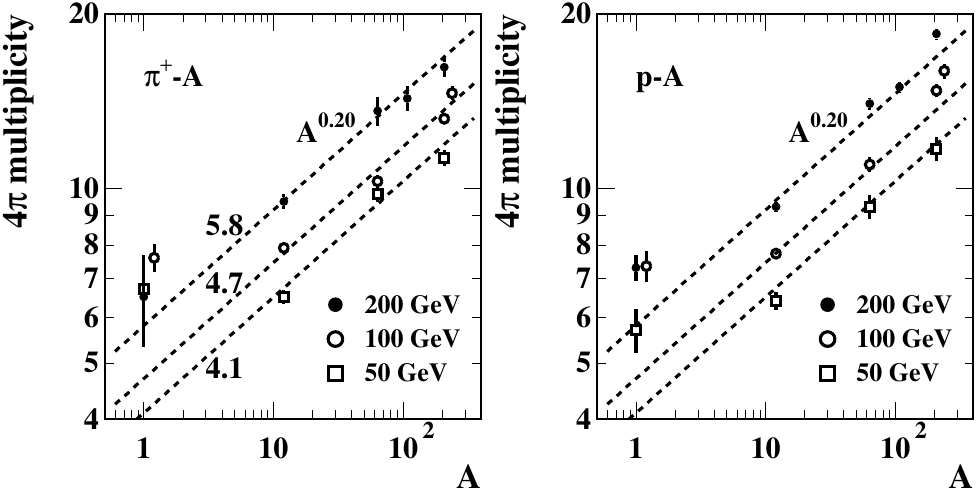}	
	\caption{\label{24}
 	Total ($\approx 4\pi$) multiplicities $\bar n_A$ vs target size A from \pa\ collisions as reported by Ref.~\cite{busza4} for $\pi^+$ and proton projectiles and for 100, 200 and 300 GeV beam energies. Dashed power-law trends are the same in left and right panels.
	}  
\end{figure}

Figure~\ref{24} shows total (approximately $4\pi$) multiplicities $\bar n_A$ vs target A for three collision energies and for $\pi^+$ (left) and proton (right) projectiles  taken from Table 5 of Ref.~\cite{busza4}. The data trends are all consistent with $\bar n_A \propto A^{0.20}$ except for the hydrogen target. The dashed lines in the two  panels are identical.

These $4\pi$ particle yields are dominated by the soft component and are thus comparable to $\bar \rho_s \propto A^{0.20}$ trends from the present TCM study of C-P spectra appearing in Fig.~\ref{trends}. In Fig.~14 of Ref.~\cite{busza4} appears ratio $R_A = \bar n_A / \bar n_H$ plotted vs quantity $\bar \nu \sim A^{1/3}$. The data are fitted with function $R_1 A^a$. The fit results in its Table 12 (b) are consistent with $a \approx 0.21$ and thus with  the C-P results.

\section{$\bf p$-A density trends: C-P vs LHC} \label{lhctrends}

As noted in connection with Fig.~\ref{trends} (right), the A dependence of C-P spectra may be better understood based on more-recent studies of \pp\ and \pa\ collisions at the LHC, in particular determination of collision ``geometry'' and its relation to particle-production trends. ``Geometry'' appears in quotes because the nominal purpose of certain analysis methods is to relate collision centrality or impact parameter $b$ to an observed quantity. However, for at least some conventional centrality methods  that purpose is not achieved. The term centrality as used below then refers to determination of parameters $N_{part} = N_{bin} + 1$ (for \pa\ collisions) in relation to hadron production mechanisms. Some alternative centrality determination methods are compared in  Ref.~\cite{tomglauber}.

\subsection{Conventional collision geometry}

``Collision geometry'' for \pa\ collisions implies statistical determination over an event ensemble of the projectile impact parameter $b$, the number of nucleon participants and some measure of resulting hadron production. Relevant quantities are then $b$, $N_{part}$ and hadron density $\bar \rho_0$ respectively. Conventional approaches employ a classical Glauber Monte Carlo, {\em assuming the eikonal approximation}, to determine a relation between $b$ and $N_{part}$. It may be further assumed that an event frequency distribution on density $\bar \rho_0$ is equivalent to a differential cross-section distribution $(1/\sigma_0)d\sigma/d\bar \rho_0$ with $d\sigma \sim db^2$, thus relating $b$ to $\bar \rho_0$. An explicit hadron production model may be invoked to relate $N_{part}$ to $\bar \rho_0$ (e.g.\ simple proportionality), thus closing the circle. It was demonstrated in Ref.~\cite{tomglauber} that such approaches are problematic for \pa\ collisions.

An example may be found in  Ref.~\cite{aliceglauber} that combines a \ppb\ Glauber Monte Carlo with assumed proportionality between yield $n_x$ in a detector element (V0A) and $N_{part}$ and with a particle-production model based on the negative binomial distribution (NBD). The \ppb\ inelastic cross section is given as $\sigma_0 = 2.1\pm0.1$ b consistent with maximum impact parameter  $b_0 \approx 8.2$ fm. For the midcentral 40-60\% centrality bin $b = 5.57$ fm, $N_{part} = 7.4$ and $N_{bin} = 6.4$. $b \approx 0.7 b_0 \approx 5.7$ fm is expected for 50\% central. For 0-100\% central $N_{part} = 7.9$ and $N_{bin} = 6.9$ which should correspond to C-P minimum-bias data. Those  results may be compared to the next subsection.

\subsection{TCM collision geometry} \label{tcmgeom}

An alternative approach is based on the TCM in combination with ensemble-mean \mmpt\ data that quantitatively determine \pa\ jet production and therefore the number of \nn\ binary collisions $N_{bin}$. The relation between $N_{part}$ and $\bar \rho_0$ then transitions from a questionable assumption to a quantitative basis for \pa\ collision ``geometry.''

\begin{figure}[h]
	\includegraphics[width=1.65in]{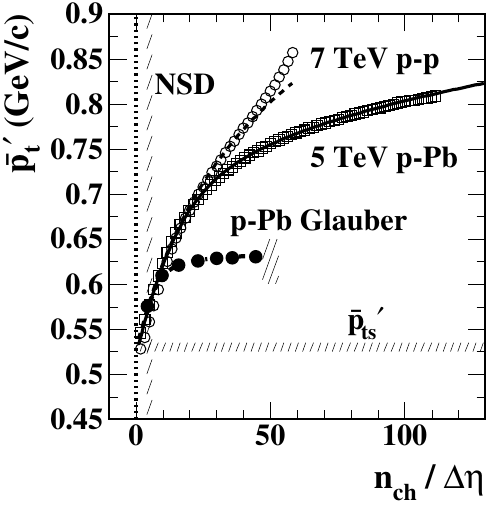}
	\includegraphics[width=1.65in]{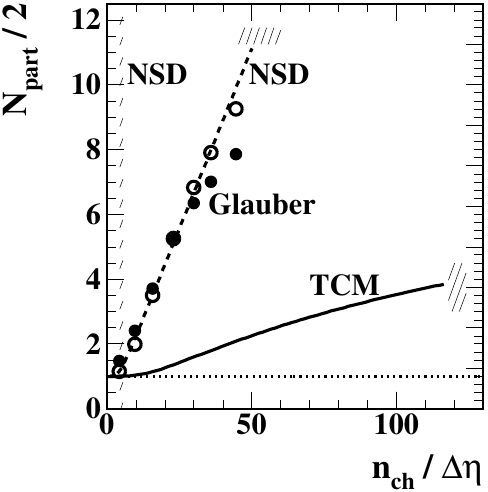}
	\caption{\label{mpt}
		Left: TCM analysis, as reported in Ref.~\cite{tommpt}, of ensemble-mean \mmpt\ data from 5 TeV \ppb\ collisions (solid curve, open boxes) and 7 TeV \pp\ collisions (dashed curve, open circles) as described in Ref.~\cite{alicempt}. The solid points and dash-dotted curve correspond to a conventional Glauber-model analysis reported in Ref.~\cite{aliceglauber}.
		Right: The relation between $N_{part}/2$ and particle density $\bar \rho_0 = n_{ch} / \Delta \eta$ according to the TCM reported in Ref.~\cite{tommpt} (solid curve) and the classical Glauber model (solid points) as reported in Ref.~\cite{aliceglauber}. The dashed line is a simple proportionality $ N_{part}/2 \approx\bar \rho_0 / 4.5$.
	}  
\end{figure}

Figure~\ref{mpt} (left) shows uncorrected (for low-\pt\ spectrum cutoff) ensemble-mean \mmpt\ data (boxes) from Ref.~\cite{alicempt}. The solid curve is a TCM description based on determining a simple relation between $N_{part}$ and $\bar \rho_0$ reported in Ref.~\cite{tommpt}. The open circles are \mmpt\ data from 7 TeV \pp\ collisions also reported in Ref.~\cite{alicempt} with a TCM description (dashed). The solid dots and dash-dotted curve are derived from data and assumptions made within a conventional Glauber-model analysis reported in Ref.~\cite{aliceglauber}.
                    
Figure~\ref{mpt} (right) shows the inferred TCM relation between $N_{part}$ and $\bar \rho_0 = n_{ch}/\Delta \eta$ (solid curve) {\em required} by \mmpt\ data. Points and dashed line represent the Glauber centrality determination reported in Ref.~\cite{aliceglauber}. Note that the Glauber description does not extend beyond $\bar \rho_0 = 50$ whereas \mmpt\ data from Ref.~\cite{alicempt} extend out to $\bar \rho_0 = 115$.

To proceed, an event frequency distribution is required within the same midrapidity acceptance used for  Ref.~\cite{alicempt} \pa\ \mmpt\ data. In the absence of published frequency data published statistical uncertainties for the \mmpt\ data can be used to construct a normalized event frequency distribution $P(n_{ch})$ on $\bar \rho_0$ as described in Ref.~\cite{tomglauber}. 

\begin{figure}[h]
	\includegraphics[width=1.67in]{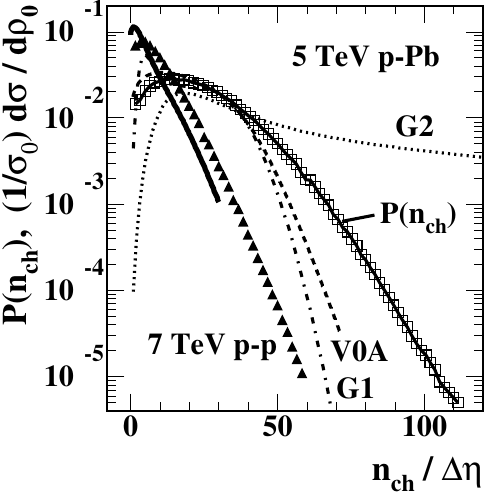}
	\includegraphics[width=1.67in]{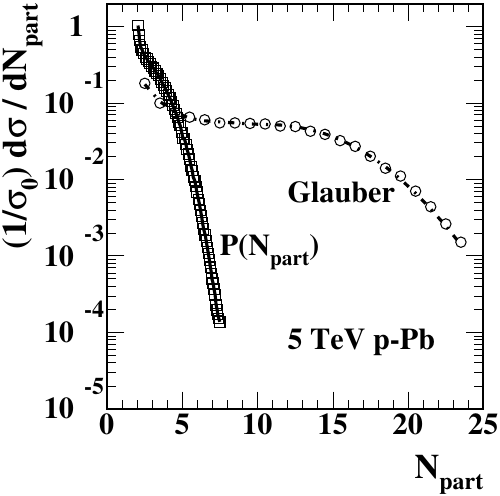}
	\caption{\label{pxx}
		Left: Event frequency distribution $P(n_{ch})$ derived from statistical uncertainties for \ppb\ \mmpt\ data in Ref.~\cite{tomglauber} (solid curve, open boxes) and from a V0A detector element of the ALICE detector as reported in Ref.~\cite{aliceglauber} (dashed). The triangles are derived from statistical uncertainties for  7 TeV \pp\ \mmpt\ data in Fig.~\ref{mpt} (left). The bold solid curve is a double-NBD fit to a probability distribution on \nch\ for 7 TeV \pp\ collisions~\cite{aliceppmult}. Glauber curves G1 and G2 are described in the text. 
		Right: Differential cross section $(1/\sigma_0)d\sigma/ dN_{part}$ derived from a Glauber Monte Carlo as described in Ref.~\cite{aliceglauber} (dash-dotted curve, open circles) and  event-frequency distribution  $P(n_{ch})$ transformed from the left to a density on $N_{part}$ via  Jacobian corresponding to  solid curve in Fig.~\ref{mpt} (right).
	} 
\end{figure}

Figure~\ref{pxx} (left) shows  frequency distribution $P(n_{ch})$ inferred from \mmpt\ statistical uncertainties (open boxes, solid curve). Also shown is a frequency distribution from \mbox{ALICE} detector element V0A (dashed curve) reported in Ref.~\cite{aliceglauber} as a density on some variable $n_x$ transformed here to a density on $\bar \rho_0$ with an appropriate Jacobian. Distribution $P(n_{ch})$ is broader than the V0A data because the jet contribution that broadens such distributions~\cite{ppquad} is substantially greater near midrapidity $\eta \in [-0.3,0.3]$ than within the V0A $\eta$ acceptance $\eta \in [2.8,5.1]$. The solid curve in Fig.~\ref{mpt} (right) then defines a Jacobian to transform $P(n_{ch})$ from $\bar \rho_0$ to $N_{part}$.

Note that \pp\ \mmpt\ data in Fig.~\ref{mpt} (left) extend out to $\bar \rho_0 \approx 60$. The same procedure can be applied to the statistical uncertainties of those data to obtain a \pp\ event frequency distribution (triangles at left) with mean value 9.5. The bold solid curve is a double-NBD fit to a 7 TeV NSD \pp\ distribution from Ref.~\cite{aliceppmult} with mean multiplicity 5.6. The measured 7 TeV \pp\ NSD  mean is $\bar \rho_0\approx 5.8$~\cite{nsdppp}.  The measured 5 TeV \ppb\ NSD mean density is $17.2\pm0.7$~\cite{alicensd}. The V0A (dashed) mean is 19 while the \ppb\ $P(n_{ch})$ (boxes) mean is 22.

Figure~\ref{pxx} (right) shows the \mmpt\ $P(n_{ch})$ (boxes) transformed from the left panel via the solid curve in Fig.~\ref{mpt} (right). Also shown is the result of a Glauber Monte Carlo (dash-dotted curve, circles) reported in Ref.~\cite{aliceglauber}. Dash-dotted and dotted curves in the left panel are the Glauber density on $N_{part}$ at right reverse transformed to densities on $\bar \rho_0$ via a constant Jacobian (G1) as assumed in Ref.~\cite{aliceglauber} (i.e.\ $N_{part} \propto n_{ch}$) and the TCM Jacobian (G2) derived from  the solid curve in Fig.~\ref{mpt} (right) based on \mmpt\ data. Note that G1 corresponds approximately to the V0A curve (dashed) up to $\bar \rho_0 \approx 50$ beyond which the Glauber representation fails entirely (see Fig.~\ref{mpt}). 

The large difference between Glauber and TCM is evident. In Ref.~\cite{tomexclude} it is demonstrated that large deviations from a classical Glauber model (assuming an eikonal approximation) may result from {\em exclusivity}: an established \pn\ interaction excludes other {\em simultaneous} interactions, thereby greatly reducing the total number of \pn\ collisions corresponding to a given \pa\ impact parameter. Exclusivity may also explain the quadratic trend $\bar \rho_h \propto \bar \rho_s^2$ observed for \pn\  (within \pa) {\em and} isolated \pp\ collisions.

\begin{figure}[h]
	\includegraphics[width=1.67in]{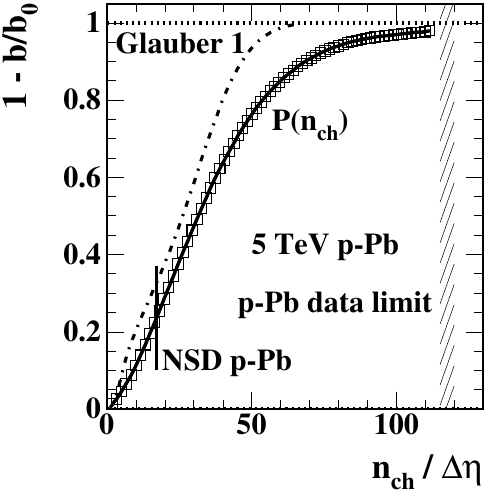}
	\includegraphics[width=1.63in]{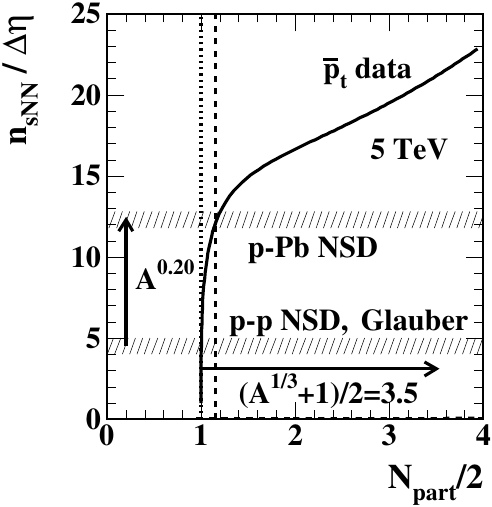}
	\caption{\label{slope}
		Left: Fractional impact parameter in the form $1 - b(n_{ch})/b_0$ for Glauber Monte Carlo (dash-dotted) and \mmpt\ uncertainties (solid curve, open boxes) assuming distributions $P(n_{ch})$ in Fig.~\ref{pxx} (left) are differential cross sections with running integrals $1 - \sigma(n_{ch})/\sigma_0$. 
		Right: Per-participant-pair density $\bar \rho_{sNN}$ vs $N_{part}/2$ (solid) from the solid curve in Fig.~\ref{mpt} (right). Hatched bands represent mean densities for NSD \pp\ (lower) and NSD \ppb\ (upper). The dashed line represents $N_{part}/2 \approx 1.15$ for \ppb\ event-class 5 ($\approx$ NSD) $\bar \rho_{sNN}$~\cite{tomglauber}.
	} 
\end{figure}

Figure~\ref{slope} (left) is based on assuming  that the density distributions in Fig.~\ref{pxx} (left) represent differential cross sections (as assumed in Ref.~\cite{aliceglauber}) in which case running integrals on $\bar \rho_0$ should yield fractional cross sections in the form $1 - \sigma(n_{ch})/\sigma_0$. Assuming $\sigma/\sigma_0 = (b/b_0)^2$ (where $b_0 \approx 8$ fm for \ppb) leads to the conjectured fractional impact-parameter trends at left. The vertical line marks NSD mean $\bar \rho_0 \approx 17$ for a 5 TeV \ppb\  event ensemble. The Glauber-1 value for $1- b/b_0$ just above 0.3 for the NSD mean is consistent with the assignment of \ppb\ event class 5 (with $\bar \rho_0 \approx 15.8$) as 40-60\% central in Ref.~\cite{aliceglauber}. However, for those values of $\bar \rho_0$ $N_{part}/2$ is substantially less than 1.5 as indicated in Fig.~\ref{mpt} (right).

Figure~\ref{slope} (right) shows TCM centrality details relevant to density trends inferred from C-P spectrum data in the present study. The solid curve is equivalent to the solid curve in Fig.~\ref{mpt} (right) but here plotted as $\bar \rho_{sNN} = (2/N_{part}) \bar \rho_s$ vs $N_{part}/2$. The hatched bands represent NSD $\bar \rho_{sNN}$ {\em averages} for \pp\ and \ppb\ collisions as follows: NSD average $\bar \rho_0 = 17.2$ corresponds to  $N_{part}/2 \approx 1.2$ from the solid curve in Fig.~\ref{mpt} (right) leading to NSD $\bar \rho_{0NN} = 17.2 / 1.2 = 14.3$. The hard component is $\bar \rho_{hNN}\approx \alpha \bar \rho_{sNN}^2 \approx 2$  leading to $\bar \rho_{sNN} \approx  12.3$. The dashed line represents $N_{part}/2 \approx 1.15$ for \ppb\ event class 5 ($\approx$ NSD)~\cite{tomglauber}. The vertical arrow, representing $A^{0.20} \rightarrow 208^{0.20} \approx 2.9$, connects $\bar \rho_s \approx 4.7$ for NSD \pp\ to 13.6 for NSD \ppb\ that reduces to $\bar \rho_{sNN} = 11.4$.  The lower arrow represents the hypothesis that $N_{bin} \sim A^{1/3}$ in which case $N_{part}/2 \approx 3.5$. Reference~\cite{aliceglauber} assumes that $\bar \rho_{0NN} \approx 4.5$ independent of $N_{part}$ (see Fig.~\ref{mpt}, right). 

\subsection{Why is NSD $\bf p$-A $\bf \bar \rho_s \propto A^{0.20}$?}

An NSD or minimum-bias density mean value is an average over an entire event ensemble such as $P(n_{ch})$ in Fig.~\ref{pxx} (left) which should be distinguished from events within a narrow ``centrality'' bin corresponding to the mean value itself. In Fig.~\ref{slope} (right) the \ppb\ NSD mean (upper hatched band) arises from the entire solid TCM curve extending up to $N_{part} / 2 \approx 4$ whereas events near the NSD mean correspond to $N_{part} / 2 \approx 1.2$ (dashed line). The lower half of $P(n_{ch})$ is dominated by single \pn\ collisions with $N_{part} / 2 \approx 1$. However, at the NSD $\bar \rho_0$ value $\bar \rho_{sNN}$ for single events (upper hatched band) has more than doubled from the \pp\ value suggesting that a second \pn\ collision within a \ppb\ event leads to substantially greater soft hadron production {\em per \pn\ collision}. The situation is dramatically different from the Glauber assumption in Ref.~\cite{aliceglauber} that all \pn\ collisions are statistically equivalent (confined to the lower hatched band in Fig.~\ref{slope}, right). The trend $\bar \rho_s \propto A^{0.20}$ thus appears to result from soft hadron production from a given participant depending on the number of prior interactions of the projectile proton which in turn depends on target A.

It is important to acknowledge the dramatic result of Fig.~\ref{slope} (right). As noted, the per-\nn\ {\em soft} density $\bar \rho_{sNN}$ at the \ppb\ NSD mean is already 2.5 times greater than the \pp\ mean which implies that the projectile proton is already strongly affected by a second peripheral \pn\ collision. The per-\nn\ value continues to increase until, for central \ppb\ collisions with $N_{part} / 2 \approx 4$, the per-\nn\ value has increased to nearly five times the \pp\ NSD value. That increase must be attributed to further modification of the projectile proton such that the average value over all \pn\ pairs follows that trend. The {\em hard} component  $\bar \rho_{hNN}$ goes as the {\em square} of the soft component and thus increases by nearly 25 times from peripheral to central \ppb\ collisions. The combination explains how the \ppb\ event distribution extends out to $\bar \rho_0 = 115$ (not 45).
For $\bar \rho_{sNN} \approx 20$,  $\bar \rho_{s} \approx 20\times 4 = 80$ and $\bar \rho_{hNN} \approx 0.013 \times 20^2 \approx 5$, so $\bar \rho_{h} \approx 7 \times 5 = 35$ with $\bar \rho_0 = 80 + 35 = 115$. It also explains how the \mmpt\ trend increases so dramatically in Fig.~\ref{mpt} (left). The \mmpt\ hard component (above  $\bar p_{ts}' \approx 0.53$ GeV/c) increases from $0.58-0.53 \approx 0.05$ GeV/c for NSD \pp\ to $0.81-0.53 \approx 0.28$ GeV/c for central \ppb, a factor five as expected. A model {\em without explicit jet production} and the quadratic relation between $\bar \rho_{hNN}$ and $\bar \rho_{sNN}$, as for instance the Glauber manifestation in Fig.~\ref{mpt}, must deviate dramatically from measured \mmpt\ and \nch\ trends.

\subsection{Why is NSD $\bf p$-A $\bf \bar \rho_h \propto A^{0.40}$?}

Given observed soft-density trend $\bar \rho_{s} \propto A^{0.20}$ the hard-density trend is easier to understand within a TCM context. Variation of $\bar \rho_{hi}$ exponent $a_{hi}$ with hadron species $i$ in Fig.~\ref{trends} (left) may be attributed to initial-state species bias at collision energies just above the threshold for jet production where valence quarks play a dominant role.  It is simpler to focus on $\bar \rho_{h} \propto A^{0.40}$  at LHC energies.

The TCM relation $\bar \rho_{hNN} \propto \bar \rho_{sNN}^2$ is fundamental  for \pn\ collisions within \pa\ collisions approximated by linear superposition~\cite{ppprd,ppbpid,tomglauber}. In the TCM description of \pa\ collisions $\bar \rho_s = (N_{part}/2) \bar \rho_{sNN}$ and $\bar \rho_h = N_{bin} \bar \rho_{hNN}$. For \ppb\ collisions near NSD mean (i.e.\  peripheral) one observes that both $N_{part}/2$ and $N_{bin}$ are greater than one by small numbers $\ll 1$. Assuming $N_{part}/ 2 \approx 1+ x$
\bea
N_{bin} &=& N_{part}-1 = 1 + 2x \approx (N_{part}/2)^2.
\eea
With that result $\bar \rho_h =  N_{bin} \bar \rho_{hNN} \propto (N_{part}/2)^2 \bar \rho_{sNN}^2 \approx \bar \rho_s^2$ for peripheral \ppb\ collisions corresponding to an NSD \pa\ average. It follows then that the trend for hard densities with target size A, at least for LHC energies, is $\bar \rho_h \propto \bar \rho_s^2 \propto A^{0.40}$ consistent with Fig.~\ref{trends} (right).

\subsection{NSD $\bf p$-A density-trends discussion}

To understand NSD \pa\ density trends one must unravel two hadron production mechanisms, the role of \nn\ fluctuations and the role of exclusivity in \pa\ collisions. An implicit assumption in Ref.~\cite{aliceglauber} appears to be a large asymmetry between \nch\ and $N_{part}$ fluctuations consistent with $N_{part} \rightarrow  n_{ch}$: i.e.\ $N_{part}$ variations dominate. Ironically, as Sec.~\ref{tcmgeom} and \mmpt\ data demonstrate, the situation is reversed: variation of particle density $\bar \rho_0$ is dominated by $\bar \rho_{0NN}$ (\nn\ density) fluctuations up to the NSD \pa\ mean over which interval $N_{part}/2 \approx 1$ is maintained. That trend is the result of large \nn\ fluctuations arising within parton splitting cascades combined with exclusivity for projectile protons within \pa\ collisions.

The phrase ``collision geometry determination'' refers to relating an impact parameter or differential cross section to an observable such as \nch, but that is apparently quite difficult and possibly irrelevant. If $N_{part}$ is accurately related to $\bar \rho_x$ (total, hard or soft densities) as in Fig.~\ref{slope} (right) then the collision dynamics may be understood. Event frequency distributions such as $P(n_{ch})$ in Fig.~\ref{pxx} (left) may be used to define event classes but do not qualify as differential cross sections directly related to impact parameter. For example, \ppb\ event class 5 defined as 40-60\% central ($\Rightarrow b/b_0 \approx 0.7$) in Ref.~\cite{aliceglauber},  consistent with a NSD \ppb\ result in Ref.~\cite{alicenucmod}, actually corresponds to $b \approx b_0$ or $N_{part}/2 \approx 1$  in Fig.~\ref{slope} (right). In Fig.~\ref{slope} (left) the $1 - b/b_0$ trend derived from  $P(n_{ch})$ (boxes) should be near zero at the NSD $\bar \rho_0$ value (vertical line) consistent with the right panel. As noted, the Glauber approach assigns the NSD mean to 40-60\% central or $1 - b/b_0 \approx 0.3$. The plotted distributions clearly do not qualify as representing fractional cross sections.

\section{C-P PID Spectrum Ratios} \label{cpratios}

The C-P spectrum analyses reported in Refs.~\cite{cronin10} (PRD11) (Figs.~9, 11 and 13) and \cite{cronin0} (PRD19) (Figs.~5, 6 and 14) include spectrum ratios (shown in  noted figures) comparing particles to antiparticles and species to species. Such ratios may reveal dependences on initial-state isospin, collision-energy and A dependences of spectra for several hadron species. In this section a  detailed study of antiparticle/particle ratios, species/species ratios and A/B spectrum ratios for C-P spectra is presented, including new details inferred from the TCM.

\subsection{Antiparticle/particle ratios for two energies} \label{antiratios}

Figure~13 of Ref.~\cite{cronin10} shows particle/antiparticle ratios for pions, charged kaons and protons from \pw\ collisions at three energies plotted vs \pt.  In that figure it is suggested that ratios vary linearly, generally increasing with \pt\ and more so with greater hadron mass. Figure~5 (left) of Ref.~\cite{cronin0} shows particle/antiparticle ratios for pions from \pp\ collisions at three energies plotted vs $x_\perp = 2 p_\perp / \sqrt{s}$. It is noted that the ratio increases from near one at low \pt\ to 2.5 or more at high \pt. One may question whether that format is most easily interpretable.

\begin{figure}[h]
	\includegraphics[height=1.65in]{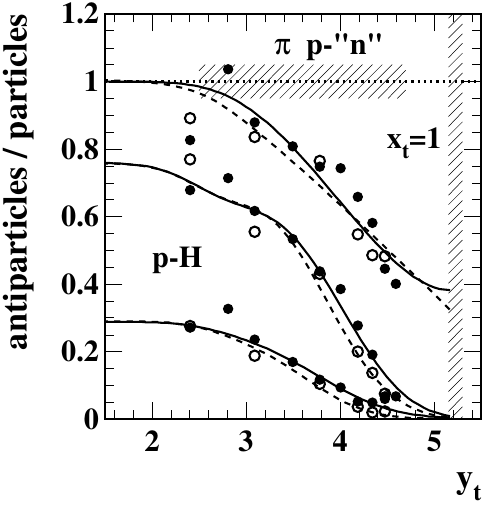}	
 \includegraphics[height=1.65in]{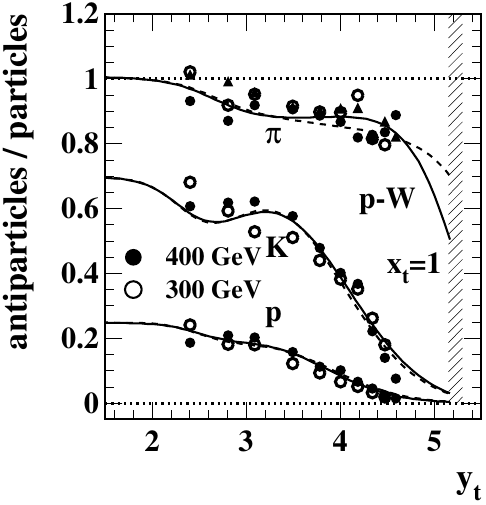}
	\caption{\label{50a}
		Left: Antiparticle/particle ratios for pions, kaons and protons from \ph\ collisions as reported in Ref.~\cite{cronin0} and for 400 GeV (solid dots, solid curves) and 300 GeV (open circles, dashed curves) beam energies. The horizontal hatched band summarizes results for $p$-``$n$'' data (see text).
		Right: Same for spectra from \pw\ collisions with data from Ref.~\cite{cronin10} except solid triangles for pions from Ref.~\cite{cronin0}. The last confirms that any possible efficiency differences between data in Ref.~\cite{cronin0} and Ref.~\cite{cronin10} cancel in antiparticle/particle ratios. The hatched bands correspond to $x_t = 1$ for 400 GeV beams.
	}  
\end{figure}

Figure~\ref{50a} (left) shows antiparticle/particle ratios (i.e.\ {\em inverse} of C-P ratios) from \pp\ collisions for 400 GeV (solid dots) and 300 GeV (open circles) plotted vs pion transverse rapidity \yt. The data trends so extrapolated indicate a possible {\em effective cutoff} for kaons and protons near \yt\ = 4.7, a cutoff not at all evident in the plotting format of Ref.~\cite{cronin10}.
 The solid (400 GeV) and dashed (300 GeV) curves are derived from TCM \pp\ spectrum trends shown in Sec.~\ref{speceng}, without adjustment.
Increase of the \pp\ pion particle/antiparticle ratio to greater than 2 noted in Ref.~\cite{cronin10} can be understood in the present format as the pion {\em antiparticle/particle} ratio descending toward a cutoff of its own.  For \pp\ data the pion ratio drops steeply towards a possible cutoff near \yt\ = 5 ($p_t \approx 10$ GeV/c).  These  trends are not unexpected: the greater the net charge of the collision system the more the opposite charge is suppressed, reflecting details at the parton level. The horizontal hatched band shows $p$+``$n$'' results from Fig.~5 (b) of Ref.~\cite{cronin0} obtained from $p$-$d$ spectra by subtraction that indicate no suppression.

Figure~\ref{50a} (right) shows antiparticle/particle ratios  from \pw\ collisions for 400 GeV (solid dots) and 300 GeV (open circles). The curves are again derived from TCM results presented in Sec.~\ref{speceng}. A major difference from the left panel is the abundance of $\pi^-$ at higher \pt\ for \pw\ collisions. The remaining deviation from unity suggests that \pw\ collisions are still peripheral as indicated in Sec.~\ref{tcmgeom}. That result corresponds to the major difference between pion ratios in Fig.~13 of Ref.~\cite{cronin10} and in Fig.~5 (left) of Ref.~\cite{cronin0}: $\pi^-$ are strongly suppressed in the smaller \pp\ collision system but not in \pw. Note that solid dots and open circles are data from Ref.~\cite{cronin10}, but the inefficiency for PRD11 data cancels in the ratio. The solid triangles represent \pw\ {\em ratios} from Ref.~\cite{cronin0} that agree with Ref.~\cite{cronin10} within uncertainties.

\subsection{Species/species ratios for targets H$\bf _2$ and W}

Ratios of differential cross sections for charged kaons and protons to pions are reported for \pw\ collisions in Ref.~\cite{cronin10} (Tables II and III and Figs.~9 and 11) and \pp\ collisions in Ref.~\cite{cronin0} (Tables IV and V and Fig.~6). Reference~\cite{cronin0} points out that the ratios are more precise than individual cross sections that include substantial normalization uncertainties. In this subsection spectrum ratio data are compared with TCM equivalents. Such comparisons provide explanations for spectrum ratio trends.

Based on Eq.~(\ref{crontcm}) TCM spectrum ratios relating hadron species $i$ to species $j$  can be expressed as
\bea \label{ijrat}
R_{ij} &=& \frac{\bar \rho_{si}(A) \hat S_{0i}(p_t,\sqrt{s}) +   \bar \rho_{hi}(A) \hat H_{0i}(p_t,\sqrt{s})}{\bar \rho_{sj}(A) \hat S_{0j}(p_t,\sqrt{s}) +   \bar \rho_{hj}(A) \hat H_{0j}(p_t,\sqrt{s})}~~
\\ \nonumber 
&\rightarrow& \frac{\bar \rho_{si}(A) \hat S_{0i}(p_t,\sqrt{s})}{\bar \rho_{sj}(A) \hat S_{0j}(p_t,\sqrt{s})} ~~\text{for low \pt}
\\ \nonumber 
\eea
In the case of hadron species $i$ and $j$ with different masses soft component model functions $\hat S_{0i}(p_t,\sqrt{s})$ do not cancel in the low-\pt\ ratio limit. However, charge densities $\bar \rho_{si}(A)$ have the same target dependence $\propto A^{0.20}$ (see Fig.~\ref{trends}), and that factor should cancel leaving a low-\pt\ limiting form approximately independent of A.

\begin{figure}[h]
	\includegraphics[width=1.65in]{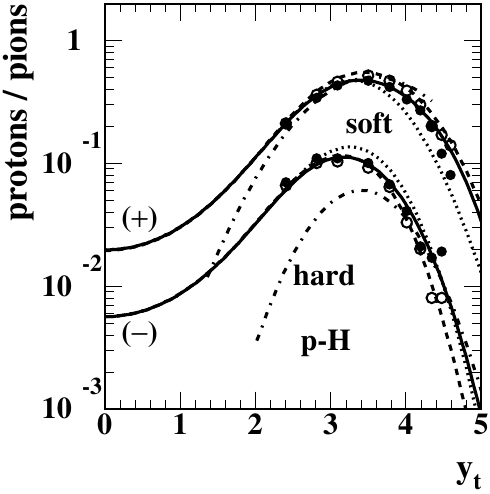}
	\includegraphics[width=1.65in]{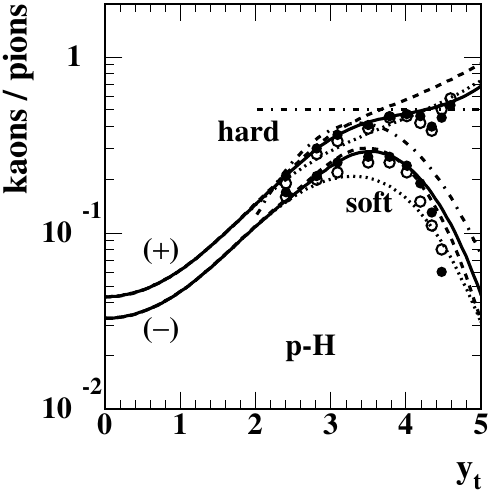}
	\put(-143,105) {\bf (a)}
	\put(-23,105) {\bf (b)}\\
	\includegraphics[width=3.4in]{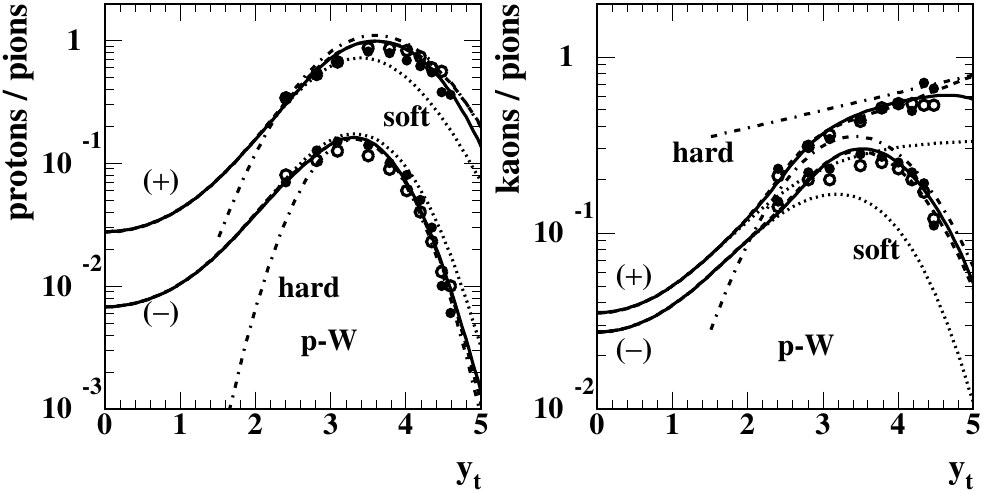}
	\put(-141,112) {\bf (c)}
	\put(-21,110) {\bf (d)}
	\caption{\label{n0x}
		Species/species spectrum ratios for protons/pions (a,c) and kaons/pions (b,d). Solid and open points are for 400 and 300 GeV beam energies, as are  TCM solid and dashed curves. Dash-dotted and dotted curves are for TCM hard and soft components. Particle and antiparticle ratios are denoted by $(+)$ and $(-)$.
	} 
\end{figure}

Figure~\ref{n0x} shows spectrum ratios for two hadron species (vs $\pi$) and two collision systems. TCM solid and dashed curves represent 400 and 300 GeV beam energies. Dotted and dash-dotted curves represent 400 GeV TCM soft and hard components. TCM curves in (a) and (b) correspond to 300 and 400 GeV spectrum curves in Fig.~\ref{20aa}. Curves in (c) and (d) correspond to 300 and 400 GeV curves in Figs.~\ref{20bb}, \ref{22bb} and \ref{21bb} (left), all without adjustment.

Trends at lower \pt\ have approximately the same shape for \ph\ vs \pw\ collisions (no A dependence), and for particles vs antiparticles (no mass variation). Note that {\em soft}-component models include a power-law tail (required for any data) whereas C-P data do not require an exponential (on \yt) tail for {\em hard} components. Thus, at some high \pt\ the soft-component tail may exceed the falling hard-component Gaussian as noted above. Trends at higher \yt\ depend on jet production in a given collision system and the relative abundance of antiparticles vs particles. 

For protons in left panels, data \yt\ trends are dominated by ratios of soft components $\hat S_0(p_t)$. Given the mass difference the proton unit-normal soft component is ``boosted'' to higher \yt\ relative  to the pion soft component such that it falls well below the pion model at lower \yt\ but rises well above it at higher \yt. However, the proton power-law tail decreases more rapidly than the pion tail (see $n$ values in Table~\ref{2aanparamsq}) and the ratio descends rapidly at higher \yt\ beyond the peak. For protons the hard component is well above the soft component near and above the peak, and substantially more so for \pw\ than for \pp. For {\em antiprotons} the hard component  falls below the soft component. Widths of baryon hard components are substantially smaller than for mesons explaining  steep hard-component ratio falloffs at higher \pt.

Trends for kaons are quite different. The smaller mass difference is manifested in a reduced soft-component ratio excursion from low \yt\ to $y_t \approx 3$. For  $K^+$ the similarity of pion and kaon $n$ values in Table~\ref{2aanparamsq} is manifested in soft-component ratios {\em not} falling off at higher \yt. $K^+$ hard-component ratios are straight lines indicating similar Gaussian widths for the two meson species. Nonzero ratio slopes correspond to slightly displaced hard-component  modes for kaons and pions. Those trends explain why $K^+$ spectrum ratios do not fall off at higher \pt. On the other hand, $K^-$ hard-component widths are substantially reduced (see Table~\ref{pwwidthsxz}) leading to strong fall-off of hard-component and total-spectrum ratios for $K^-$. The complex trends exhibited by species/species ratios as described here can only be understood by means of differential TCM analysis described in Secs.~\ref{speceng} and \ref{cpadep}.

\subsection{A/B spectrum ratios and NMFs} \label{nmfs}

Spectrum ratios comparing target A to target B (as for \pa\ spectra from Sec.~\ref{cpadep}) are conventionally referred to as nuclear modification factors (NMFs) in relation to possible jet modification or ``quenching'' as the result of a dense medium. {\em Rescaled} spectrum ratios are variously defined in that context but, as noted above, a common example based on differential cross sections is~\cite{accardi}
\bea \label{cronratio}
R_{AB}(p_t) &=& (A/B)^{-1}\frac{Ed\sigma_{pA}/d^3p}{Ed\sigma_{pB}/d^3p},
\eea
where cross sections $Ed\sigma_{pX}/d^3p$ are defined in terms of some $p$-A minimum-bias cross section $\sigma_{pA}$ as a factor. According to Ref.~\cite{accardi} ``In absence of nuclear effects one would expect $R_{AB} = 1$, but for $A>B$ a suppression is observed experimentally at small \pt, and an enhancement at moderate \pt\ with $R_{AB} \rightarrow 1$ as $p_t \rightarrow \infty$.'' Several assumptions appear to underly such expectations: (a) $N_{bin} \approx A^{1/3} \sim A \sigma_{pN}/\sigma_{pA}$ (also denoted by $\bar \nu$) is an estimated number of \nn\ binary collisions leading to jet production. (b) Hadron densities associated with jet production do not vary significantly with target A. (c) Spectrum structure associated with jet production does not vary within a conventional QCD context. Those assumptions are probably incorrect per the material in Sec.~\ref{lhctrends}.

Unrescaled particle {\em density} spectrum ratios can be expressed via the TCM as 
\bea 
\label{abratp}
R'_{AB}(p_t) &=& \frac{\bar \rho_{si}(A) \hat S_{0i}(p_t,\sqrt{s}) +   \bar \rho_{hi}(A) \hat H_{0i}(p_t,\sqrt{s})}
{\bar \rho_{si}(B) \hat S_{0i}(p_t,\sqrt{s}) +   \bar \rho_{hi}(B) \hat H_{0i}(p_t,\sqrt{s})}~~~~
\textit{}\\ \nonumber 
&\rightarrow& \frac{\bar \rho_{si}(A) }
{\bar \rho_{si}(B)} \approx (A/B)^{a_s} ~~\text{for low \pt}
\\ \nonumber 
&\rightarrow& \frac{\bar \rho_{hi}(A) }
{\bar \rho_{hi}(B)} \approx (A/B)^{a_h} ~~\text{for high \pt}.
\\ \nonumber 
\eea
The values of $a_s$ and $a_h$ are derived from Fig.~\ref{trends} with values shown in Fig.~\ref{power} (d).
Since factor $\sigma_{pA}(A) \propto A^{m}$ (where $m \approx 2/3$) is included in differential cross sections, ratio $R_{AB}(p_t)$ may be re-expressed in terms of unrescaled ratios of particle-density spectra $R'_{AB}(p_t)$ as
\bea \label{abrat}
R_{AB}(p_t) &\approx& (A/B)^{m-1}R'_{AB}(p_t)
\\ \nonumber 
&\rightarrow&  (A/B)^{a_s + m-1} ~~\text{for low \pt}
\\ \nonumber 
&\rightarrow&  (A/B)^{a_h + m-1} ~~\text{for high \pt}
\\ \nonumber 
\eea
The effective value of $m$ depends on what $\sigma_{pA}$ cross sections are actually included.

Those relations explain Fig.~12 of Ref.~\cite{cronin0} that shows the $\pi^-$ differential cross-section spectrum for \pw\ (not rescaled and including factor $\sigma_{pA}$) vs atomic weight A in a log-log format using spectrum values at $p_t = 4.6$ GeV/c that are thus dominated by the hard component. Assuming $m \approx 2/3$ the trend corresponding to Eq.~(\ref{abratp}) is $A^{2/3} \times A^{0.50} \approx A^{1.17}$ consistent with the straight line $A^{1.165}$ in Ref.~\cite{cronin0} even though $a_h \approx 0.50$ (specifically for $\pi^-$) is inferred from TCM $\bar \rho_h$ trends measured near 1 GeV/c ($y_t \approx 2.7$, see Fig.~\ref{20dd} right) while Fig.~12 of Ref.~\cite{cronin0} applies to a data trend at 4.6 GeV/c ($y_t \approx $ 4.2).

The ratio measure defined by Eq.~(\ref{cronratio}) is implicitly based on the assumption that $A^{1/3}$ may serve as an estimator for number of binary \pn\ collisions $N_{bin}$ in \pa\ collisions, in which case ``absence of nuclear effects'' in Ref.~\cite{accardi} translates to the assumption that jet production $\propto$ estimated $N_{bin}$ unless jet formation is modified (``nuclear effects''). But jet production measured by particle densities associated with spectrum hard components is the product of $N_{bin}$ and jet production from \nn\ pairs represented by $\bar \rho_{hNN}$ (see Eq.~(\ref{pidspectcm})). In \pp\ collisions the latter varies {\em quadratically} with the soft component and $N_{bin}\equiv 1$. In \pa\ collisions one finds that $\bar \rho_{sNN}$ and $\bar \rho_{hNN}$ vary strongly while $N_{bin}$ increases quite slowly. In \aa\ collisions the reverse may be the case, but it is not correct to assume  $\bar \rho_{xNN}$ is fixed independent of event class.

\begin{figure}[h]
	\includegraphics[width=3.3in]{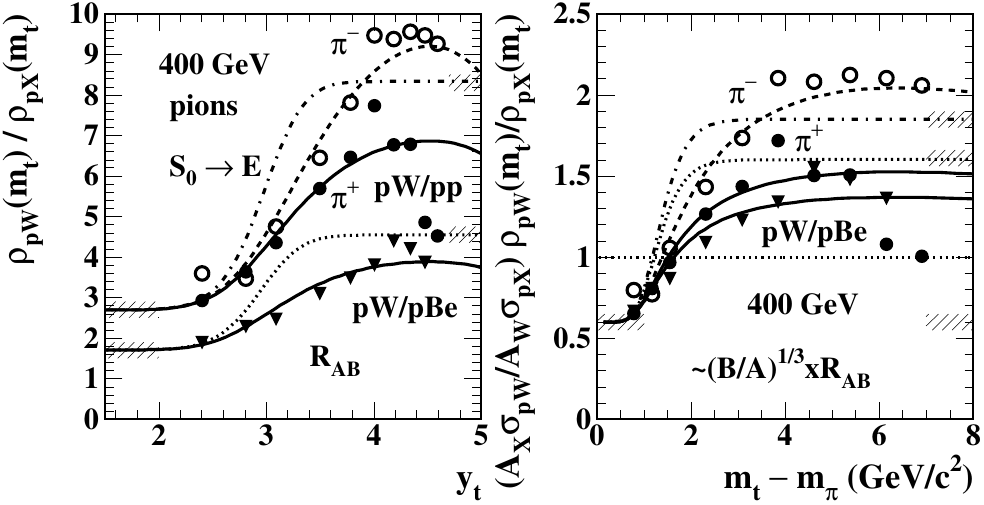}	
	\caption{\label{20ix}
		Left: Pion spectrum (particle density) ratios denoted by pW/pp and pW/pBe in the form of $R'_\text{AB}$ as defined by Eq.~(\ref{abratp}) from fixed-target energy 400 GeV \pa\ collisions for data (points) and TCM (solid and dashed curves). The data spectra are those shown in Fig.~\ref{20dd}. The dotted and dash-dotted reference curves are described in the text.
		Right: Spectrum ratios in the left panel combined with ratios of factors $\sigma_\text{$p$A}/A$ to form ratios $R_\text{AB}$ as defined by Eq.~(\ref{abrat}). 
	} 
\end{figure}

Figure~\ref{20ix} (left) shows $\pi^+$ (solid points, solid curves) and $\pi^-$ (open points, dashed curve) particle-density spectrum ratios (points) for collision systems $p$-X with {\em no rescaling}, derived from spectra reported by the C-P collaboration~\cite{cronin0}, based on ratio $R_{pA}'$ as defined by Eq.~(\ref{abratp}). Solid and dashed curves represent a TCM for spectrum data as described above which {\em assumes no A dependence} for model functions. The difference between $\pi^+$ and $\pi^-$ \pw/\pp\ ratios arises because of the large pion antiparticle/particle asymmetry for \pp\ spectra  in Fig.~\ref{50a} (left) which greatly exceeds that for \pw\ spectra (right). 

Dash-dotted (pW/pp) and dotted (pW/pBe) curves correspond to Eq.~(\ref{abratp}) with exponent $n$ in model $\hat S_0(p_t)$ set to $1/n \rightarrow 0$, i.e.\ the model function takes on a Boltzmann exponential limit and becomes negligible {\em relative to the hard component} above $y_t \approx 3$. Hatched bands show low- and high-\pt\ limiting cases for Eq.~(\ref{abratp}) determined from pion $\bar \rho_s$ and $\bar \rho_h$ values in Fig.~\ref{trends} and Table~\ref{pionaparams} that agree with the dash-dotted and dotted curve endpoints. 
Those limiting cases illustrate a basic feature of such ratio plots. The TCM soft components required to describe \mbox{C-P} data deviate strongly from a Boltzmann exponential as in Fig.~\ref{20bb} (a) (dash-dotted curve E). As a result, ratios pW/pX rise to a peak near \yt\ = 4 (above the hard-component $\bar y_t$) but then fall again because the hard-component Gaussian (with no power-law tail) drops off rapidly while the soft component continues its power-law trend at higher \pt. Factors $\hat S_{0i}(p_t)$ are not negligible at higher \pt\ compared to hard components because (a)  soft components include substantial power-law tails ($n \sim 25$) while (b) jet contributions to C-P spectra for $\sqrt{s} \sim 25$ GeV, near the jet threshold, are relatively small for pions. 

Figure~\ref{20ix} (right) shows the same data and curves rescaled by conventional factors $\sigma_{pA}/A \sim A^{-1/3}$ per Eq.~(\ref{abrat}). The $x$ axis is transverse mass \mt\ that minimizes visual access to the transition region from low-\pt\ ``soft'' processes to high-\pt\ jet phenomena. The dash-dotted and dotted curves show what to expect if the jet structure were indeed isolated and no ``nuclear effects'' were involved -- i.e.\ the hard-component {\em shape} is then independent of A and cancels in the Eq.~(\ref{abrat}) ratio. If the \ph\ cross section were increased from 40 mb (off the $A^{2/3}$ trend) to 45 mb the pW/pp and pW/pBe trends would coincide, further complicating ratio interpretations.

Low-\pt\ ``suppression'' and {\em reduced} high-\pt\ enhancement are then evident but the mechanism(s) may not be derived from spectrum ratios that discard much information carried by isolated spectra. In contrast, knowledge of density trends in Fig.~\ref{trends} allow quantitative predictions of {\em limiting values} as in Eq.~(\ref{abrat}), but deviations from those values, as in Fig.~\ref{20ix} (right), can only be understood in terms of TCM soft components identified in Sec.~\ref{ppsoft}.

In  this ratio format the jet contribution is not  properly represented. Spectrum {\em ratios} demand precision for a vanishing fraction of jet fragments. The great majority of jet fragments appears near 1 GeV/c ($y_t \approx 2.7$) in Fig.~\ref{20d}. The trends above 4 GeV/c ($y_t \approx 4$) represent {\em less than two percent} of jet fragments and give a misleading impression of any data-model disagreement. Figure~\ref{20ix} (right) also demonstrates that $R_\text{AB}$ trends at higher \pt\ have no requirement to descend to unity. Given the effect of soft-component power-law tails the ratio trend would go asymptotically to the soft-component density ratio indicated by the hatched band at lower right, albeit at a  \pt\  beyond the kinematic limit of  the C-P collision energy.

Referring to the statement of Ref.~\cite{accardi} -- ``In absence of nuclear effects one would expect $R_{AB} = 1$ [at higher \pt]'' -- that would require $\bar \rho_h(A)$ to vary as $A^{1/3}$ given the structure of Eq.~(\ref{cronratio}) and no soft-component tail contribution at higher \pt. Figure~\ref{trends} indicates that real collisions substantially deviate from those conditions in a manner consistent across a large A interval. One may then question what ``nuclear effects'' means and in what context. Such rescaled spectrum ratios may be explained technically in a TCM context but are generally misleading.

\subsection{Power-law trends $\bf A^{\alpha(p_t)}$ vs $\bf p_t$} \label{apttrends}

In Ref.~\cite{cronin0} (PRD19) considerable attention is given to variation of measured spectra with target atomic weight A. It is observed  that ``...although the [differential] cross sections did extrapolate as $A^\alpha$, the power $\alpha$ is a function of $p_\perp$ and...grows to be greater than 1.0 at large $p_\perp$....'' Figure~12 from Ref.~\cite{cronin0} is  mentioned in Sec.~\ref{nmfs} above and basically presents $d\sigma_A(p_t) / d\sigma_W(p_t) \propto A^{\alpha(p_t)}$ for $\pi^-$ and $p_t = 4.61$ GeV/c, with $\alpha \approx 1.17$. Figure 13 is then $\alpha(p_t)$ vs \pt\ for $\pi^+$ and $\pi^-$. Figure 15 is the equivalent of Fig.~12 but for $K/\pi$ and $p/\pi$ ratios. Figure 18 summarizes $\alpha(p_t)$ for $\pi^\pm$, $K^\pm$ and $p^\pm$. Figure~19 demonstrates no significant energy dependence for the $\alpha(p_t)$ trends, consistent with the present finding of no significant energy dependence for $\bar \rho_x(A)$ trends. In the following, C-P power-law trends are explored in the context of the TCM.

Given the expectations for A dependence in  Ref.~\cite{cronin0} and the definition of $R_{AB}(p_t)$ in Eq.~(\ref{cronratio})  one can write $R_{AB}(p_t) \approx (A/B)^{\alpha_{AB}(p_t)}$, in which case
\bea \label{rap}
\alpha_{AB}(p_t) &\approx& \log[R_{AB}(p_t))]/ \log(A/B)
\eea
provides simple access to $\alpha_{AB}(p_t)$ via spectrum ratios. In a TCM context one may define an A-dependent ratio
\bea \label{rprime}
R'(A)&\equiv&  \frac{\bar \rho_{spp}A^{a_s} \hat S_{0p}(p_t) + \bar \rho_{hpp}A^{a_h} \hat H_{0A}(p_t)}{\bar \rho_{spp} \hat S_{0p}(p_t) + \bar \rho_{hpp} \hat H_{0p}(p_t)}~~~
\eea
analogous to Eq.~(\ref{abratp}) for A-B ratios $R'_\text{AB}(p_t)$. If one assumes that $R'(A) \approx A^{a(p_t)}$ then
\bea \label{lograt}
a(p_t) &=& \log[R'(A)(p_t)] / \log(A)
\eea
and $\alpha(p_t) \approx a(p_t) + 2/3$ since $R'(A)$ is a ratio of particle densities, not differential cross sections with $\sigma_{pA} \propto A^{2/3}$. 

\begin{figure}[h]
	\includegraphics[width=1.65in]{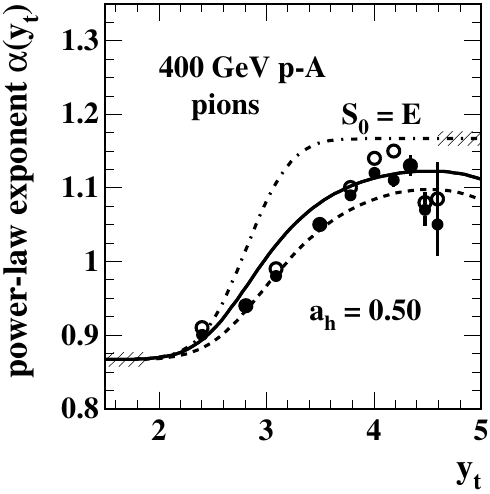}
	\includegraphics[width=1.65in]{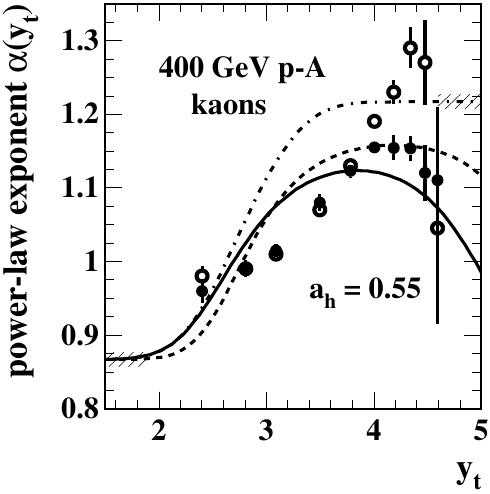}
\put(-145,30) {\bf (a)}
\put(-23,30) {\bf (b)}\\
	\includegraphics[width=1.65in]{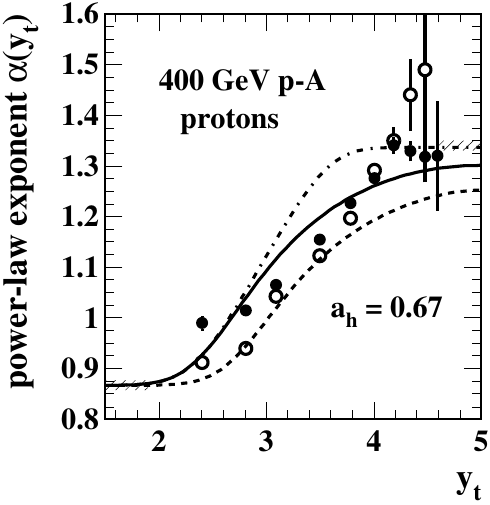}
	\includegraphics[width=1.65in]{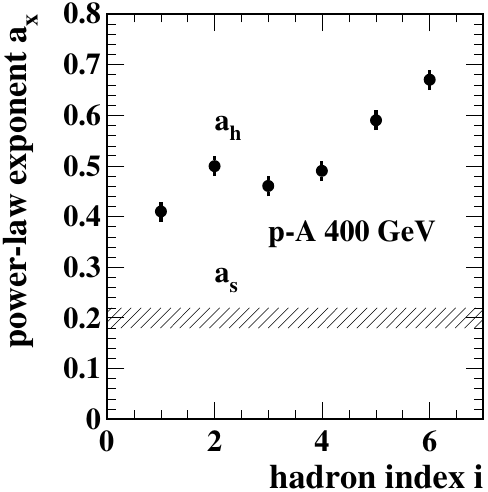}
\put(-145,30) {\bf (c)}
\put(-23,30) {\bf (d)}
	\caption{\label{power}
(a-c) $\alpha(p_t)$ data from Fig.~18 of Ref.~\cite{cronin0} for positive (solid) and negative (open) hadrons. Corresponding curves (solid and dashed respectively) represent Eqs.~(\ref{rprime}) and (\ref{lograt}) with $a_s = 0.20$ and $a_h$ values for each hadron species as noted in the panels, approximately consistent with Fig.~\ref{trends}. (d) $a_x$ values derived from detailed analysis of  trends in Fig.~\ref{trends} (left).
	} 
\end{figure}

Figure~\ref{power} (a-c) shows $\alpha(p_t)$ data presented in Fig.~18 of Ref.~\cite{cronin0} for positive (solid) and negative (open) hadrons. It may be that the $\alpha(p_t)$ data points are derived from individual fits to log-log trends as in Fig.~12 of  Ref.~\cite{cronin0}.  Solid (positive) and dashed (negative) curves show TCM trends for $\alpha(p_t)$ derived from Eq.~(\ref{rprime}) with A = 184,  fixed value $a_s = 0.20$ and with $a_h$ noted in each panel that best describe the data. The dash-dotted curves correspond to setting $\hat S_{0p}(p_t)$ to an exponential via $1/n \rightarrow 0$. The hatched bands show low- and high-\pt\ limiting cases $\alpha_s$ and $\alpha_h$ from Eq.~(\ref{lograt}) with $\alpha_x = a_x + 2/3$. The values of $\bar \rho_{xpp}$ in Eq.~(\ref{rprime}) are derived from Table~\ref{ppdensities}. For this exercise the TCM model functions remain the same in numerator and denominator of Eq.~(\ref{rprime}) [$\hat H_{0A}(p_t) \rightarrow \hat H_{0p}(p_t)$].  This figure emphasizes 400 GeV data because Fig.~19 of Ref.~\cite{cronin0} shows no significant energy dependence to $\alpha$ and the 400 GeV data have the most complete \pt\ coverage.

Figure~\ref{power} (d) shows values $\alpha_s$ and $\alpha_h$ derived from Fig.~\ref{trends} density trends comparable with results in panels (a-c). Note  that $a_h$ values from  Fig.~\ref{trends} correspond to the procedure in Sec.~\ref{atrends} such that inferred values correspond to the modes of hard components near 1 GeV/c whereas the $\alpha(p_t)$ data values in Fig.~\ref{power} (a-c) are inferred over a broad range of \pt\ values. Thus, differences between $a_h$ in panels (a-c) and panel (d) are not unexpected.

Alternatively, one could apply the approach of Eqs.~(\ref{rprime}) and (\ref{lograt}) directly to C-P data spectrum ratios $R'_{AB}(p_t)$ as in Eq.~(\ref{abratp}). If one assumes, similar to above, that $R'_\text{AB}(p_t) \approx (A/B)^{a_\text{AB}(p_t)}$ then
\bea \label{logratab}
a_\text{AB}(p_t) &=& \log[R'_\text{AB}(p_t)] / \log(A/B)
\eea
and $\alpha_\text{AB}(p_t) \approx a_\text{AB}(p_t) + 2/3$ since $R'_\text{AB}$ is a ratio of particle densities, not differential cross sections. In Eq.~(\ref{rprime}) soft component $\hat S_{0p}(p_t)$ and hard component $\hat H_{0A}(p_t)$ are assumed independent of A as noted above.

\begin{figure}[h]
	\includegraphics[width=3.3in]{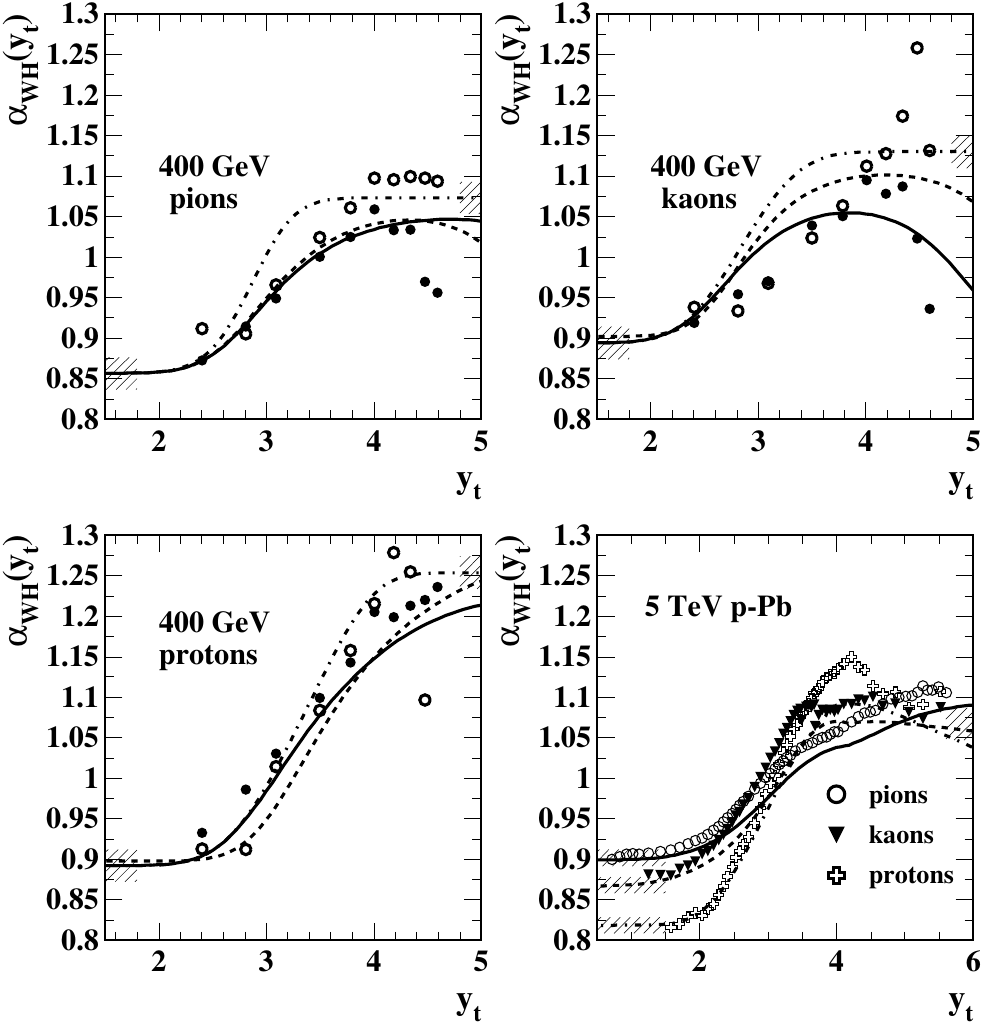}
	\caption{\label{direct} (a-c) $\alpha_\text{WH}(p_t)$ values for positive (solid) and negative (open) hadrons inferred from 400 GeV C-P spectra directly via Eqs.~(\ref{abratp}) and (\ref{logratab}). Solid and dashed curves (positive and negative hadrons respectively) are obtained in the same manner from TCM spectra that describe C-P spectra as shown in Figs.~\ref{20dd}, \ref{22dd} and \ref{21dd}. Dash-dotted curves are limiting cases described in the text. Hatched bands are inferred from soft and hard hadron densities as described in the text. (d) Results of the same procedure applied to PID spectra from 5 TeV NSD \ppb\ collisions reported in Ref.~\cite{alicenucmod}.
	} 
\end{figure}

Figure~\ref{direct} (a-c) shows $\alpha_\text{WH}(y_t)$ derived from C-P  spectrum data for pions, charged kaons and protons and for positive (solid) and negative (open) hadrons. The data spectra in the W/H ratio are those for 400 GeV \pw\ and \ph\ collisions from Ref.~\cite{cronin0} (PRD19) as they appear in Figs.~\ref{20dd}, \ref{22dd} and \ref{21dd} processed per Eqs.~(\ref{abratp}) and (\ref{lograt}) with $\alpha_\text{WH} = a_\text{WH} + 2/3$. Solid (positive) and dashed (negative) curves are derived from ratios of corresponding TCM spectra. Dash-dotted curves arise from positive-hadron TCM spectrum ratios with soft-component exponent $1/n \rightarrow 0$ as a limiting case. Hatched bands show limiting values of $\alpha_\text{WH}(y_t)$ derived from ratios $\bar \rho_x(pW) / \bar \rho_x(pH)$ with values from Tables~\ref{pionaparams} - \ref{protonaparams}.

Figure~\ref{direct} (d) shows a comparable treatment of PID spectrum data from 5 TeV \ppb\ collisions reported in Ref.~\cite{alicenucmod} where NSD \ppb\ spectra are compared in ratio to minimum-bias (MB) \pp\ spectra assuming some $N_{bin}$ estimate per the relation $R_\text{AB} \equiv (1/N_{bin})R'_\text{AB}$ (i.e.\ NMFs). The choice of NSD \ppb\ and MB \pp\ spectrum data should be compatible with minimum-bias C-P spectrum data. In Ref.~\cite{ppbnmf} a TCM study of LHC data reveals  that the NSD \ppb\ spectra are approximately equal to \ppb\ event class 5 (of 7) and that the MB \pp\ spectra are substantially biased relative to corresponding TCM event class 7 (most peripheral). In panel (d) the points are derived from ratios of NSD-averaged \ppb\ data to TCM event class 7 with spectrum ratios treated as in panels (a-c). The reference curves (solid, dashed and dash-dotted for pions, kaons and protons respectively) correspond to ratios of TCM event classes 5 and 7 where, as demonstrated in Ref.~\cite{ppbnmf}, TCM spectra are statistically equivalent to data spectra with the exception of protons for event class 7 (peripheral, not relevant in this case).

The hatched bands at low \pt\ correspond to ratios of soft-component densities $\bar \rho_{si} = z_{si}(n_s)(N_{part}/2)\bar \rho_{sNN}$ for \ppb\ event classes $n = 5$ and 7 with parameters taken from Ref.~\cite{ppbnmf}. Limiting values are then given by $\alpha = \log[\bar \rho_{si}(5)/\bar \rho_{si}(7)] / \log(208) + 2/3$. Note that  C-P spectra in panels (a-c) do not extend low enough on \pt\ to determine precise soft-component limiting cases. The hatched band at high \pt\ corresponds to $a \rightarrow a_h = 0.40$ in Eq.~(\ref{lograt}) consistent with 5 TeV results in Fig.~\ref{trends} (right) . Differences between hadron species arise solely from fractional abundances $z_{xi}(n_s)$ that depend only on hadron mass. The \ppb\ centrality or A dependence is  stronger for $z_{si}(n_s)$ than for $z_{hi}(n_s)$, consistent with results in Fig.~\ref{direct} (d) for low \pt\ vs high \pt.

Panel (d) thus presents a confusing picture of what is simple within a TCM context. The basic structure represents $a_{s} \approx 0.20$ and $a_h \approx 0.40$ from Fig.~\ref{trends} (right) combined with the assumptions associated with Eq.~(\ref{cronratio}) and the PID spectrum TCM for \ppb\ previously established in Refs.~\cite{ppbpid,pidpart1,pidpart2}. The combination is uninterpretable.

\section{Discussion} \label{disc}

Two topics are considered at greater length:
(a) theoretical conjectures about energy dependence of \pa\ spectrum soft- and hard-component shapes and
(b) comments on a contemporary review of fixed-target \pa\ experimental results that provides context for the present study.

\subsection{p-A spectrum energy dependence} \label{paedep}

Reference~\cite{cronin10} (1974) appears early in the development of QCD theory per A-B nuclear collisions. Emphasis is placed on so-called scaling variable $x_\perp  \rightarrow x_t = 2 p_t / \sqrt{s}$ in reference to predictions of the spectrum trend $\bar \rho_0(p_t) \sim g(s) f(x_t) \rightarrow e^{-ax_t}/ s^{n}$ for $x_t > 0.40$. A fit to \pw\ data as described below leads to $n \approx 5.5$ and $a \approx 36$.

\begin{figure}[h]
\includegraphics[height=1.7in]{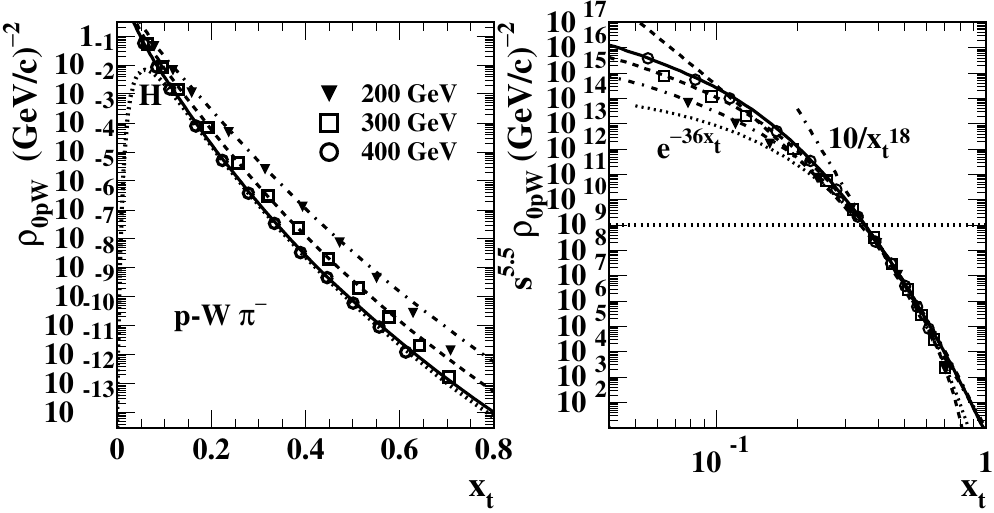}
	\caption{\label{scale}
Left: C-P $\pi^-$ \pw\ spectrum data from Ref.~\cite{cronin10} plotted in the format of its Fig.~4. Solid, dashed and dash-dotted curves are from Fig.~\ref{20bb} (c) above transformed to the same format. The bold dotted curve is the TCM hard component for 400 GeV.
Right:  C-P $\pi^-$ spectrum data from Ref.~\cite{cronin10} plotted in the format of its Fig.~6. The TCM curves are as described at left. The dotted curve is suggested to describe data in a ``scaling region.'' The upper dashed curve is a theoretical conjecture discussed further below. The upper dash-dotted curve is a power-law function introduced here as a reference.
	}  
\end{figure}

Figure~\ref{scale} (left) shows \pw\ $\pi^-$ data spectra (points) and TCM (solid, dashed and dash-dotted curves) from Fig.~\ref{20bb} plotted vs $x_t$. Also shown is 400 GeV TCM hard component $H(p_t) = \bar \rho_h\hat H_0(p_t)$ (dotted) demonstrating that $H$ alone determines C-P spectra at higher \pt\ even for this low collision-energy range. Given the conjectured  trend, quantity $s^{n}\bar \rho_0(p_t)$ is expected to exhibit ``scaling'' at higher \pt. Reference~\cite{cronin10} states that ``...at large $p_\perp$ the cross sections fall exponentially; they do not show a manifest power-law dependence.'' That conclusion appears to follow from a data description based on $e^{-ax_t}$.

Figure~\ref{scale} (right) shows $s^{n}\bar \rho_0(p_t)$ transformed from spectra in the left panel with model parameter values $n \approx 5.5$ and $a \approx 36$ as noted above. A log-log format is used to evaluate possible power-law trends in the data. The exponential trend is denoted by the dotted curve. A power-law trend is represented by the dash-dotted line. These data are not able to exclude a power-law trend for $x_t > 0.4$. The dashed curve proceeding from upper left is another theory conjecture discussed below. Dotted and dashed curves are visible at lower right whereas the dash-dotted power-law trend coincides with  TCM curves at $x_t \approx 1$. These results confirm that the TCM describes all C-P data within point-to-point uncertainties except the highest-\pt\ points for 200 and 300 GeV. The theory conjectures describe data above $x_t = 0.4$, including the two highest points noted above, but are freely fitted to the data in that interval. It is not clear how to interpret the fit parameters. In contrast, the simple TCM hard-component Gaussian on \yt\ describes data above $x_t = 0.1$. 

Referring to species/species ratios in its Figs.~9 and 11 Ref.~\cite{cronin10} states (referring to its Fig.~9) that for $x_t > 0.4$ ``...$K^+$ cross section scales in the same way as the pions,'' i.e. there is no significant variation of the ratio trend with energy. However, ``[t]he $K^-/\pi^-$ ratio...does not appear to scale....'' The $K^-$ trend is easily explained. Referring to Fig.~\ref{xxx} and Table~\ref{pwwidthsxz} above, the hard-component widths for $\pi^\pm$ and $K^+$ agree within uncertainties whereas the $K^-$ widths are substantially smaller. The same is true for protons relating to Fig.~11 of Ref.~\cite{cronin10}. In each of Figs.~9 and 11 the ratios are independent of energy for the lowest points (i.e.\ at and below the hard-component modes) but then diverge above that point due to the width differences. Increase of hard-component widths with collision energy is expected, and continues strongly up to LHC energies per underlying jet spectra~\cite{alicetomspec}. While  systematics of their Figs.~9 and 11 are easily explained as above, trends on $x_t$ in Figs.~10 and 12 are not.

Reference~\cite{cronin0} (1979)  considers subsequent theory conjectures about spectrum collision-energy and \pt\ systematics, specifically for \pp\ collisions. The suggested model is $\bar \rho_0(p_t) \sim (1-x_t)^b/p_t^n$~\cite{gunion} with fitted values $n \approx 8$ and $b \approx 9$. The conjecture is tested in its Fig.~3.

\begin{figure}[h]
\includegraphics[height=1.7in]{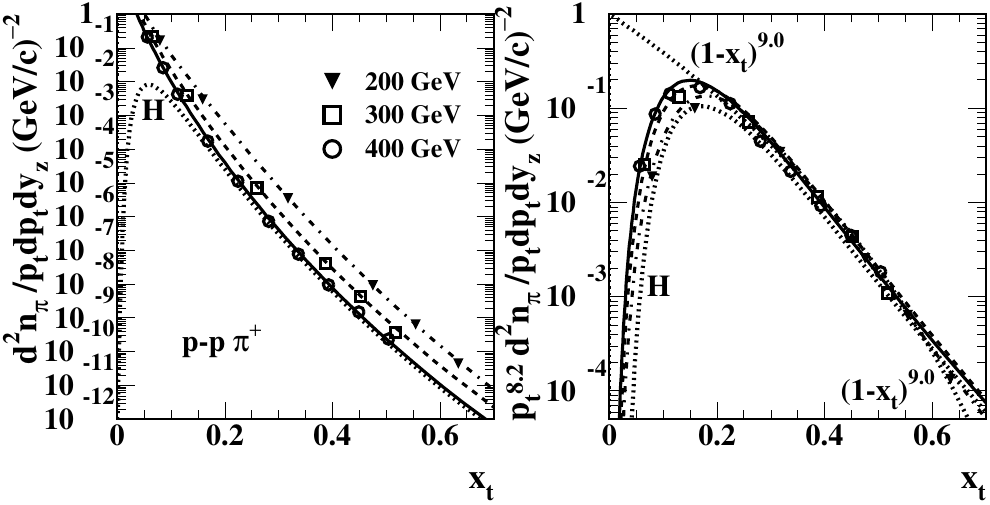}
	\caption{\label{ppscale}
Left: C-P $\pi^+$ \ph\ spectrum data from Ref.~\cite{cronin0} plotted vs parameter $x_t$ (see Fig.~\ref{scale}, left). Solid, dashed and dash-dotted curves are from Fig.~\ref{20aa} (a) above transformed to the same format. The bold dotted curve is the TCM hard component for 400 GeV.
Right:  C-P $\pi^+$ \ph\ spectrum data from Ref.~\cite{cronin0} plotted in the format of its Fig.~3 (left) including factor $p_t^{8.2}$. The upper dotted curve is function $(1-x_t)^{9.0}$ conjectured to describe data in a ``scaling region.''
	}  
\end{figure}

Figure~\ref{ppscale} (left) shows the equivalent of Fig.~\ref{scale} (left) for \pp\ collisions rather than \pw\ collisions. Solid, dashed and dash-dotted curves are the TMC from Fig.~\ref{20aa} (a). The dotted curve is hard component $H$ for  400 GeV. The TCM describes data within point-to-point uncertainties, and the hard component dominates \pp\ spectra for $x_t > 0.2$. The mode of the pion hard component as a density on \yt\ is near $y_t = 2.7$ or $p_t \approx 1$ GeV/c which corresponds to $x_t \approx 0.07$ for 400 GeV. Data-model agreement demonstrates that the spectrum data are consistent with a simple Gaussian on \yt\ above that point for C-P collision energies. As demonstrated in Sec.~\ref{speceng}, what varies with collision energy is the Gaussian width {\em above} the mode (experimentally accessible at these energies).

Figure~\ref{ppscale} (right) duplicates the left panel of Fig.~3 in Ref.~\cite{cronin0} except that the C-P data here do not include absorption cross-section factor $\sigma_{pA}$ as published. This figure confirms the ``scaling'' inferred from Fig. 3. However, relevant information carried by the TCM is  Gaussian width variation with collision energy, especially as it relates to hadron species and higher collision energies at the RHIC and LHC (see Fig.~\ref{xxx}). That information is essentially abandoned in the format of the right panel, and how are parameters $b$ and $n$ to be interpreted physically?

Reference~\cite{cronin0}  observes that ``[i]t now appears that with a theory which actually predicts the single particle spectra to be rather complicated functions of $x_\perp$ and $p_\perp$, precise measurements over a wide range in these variables will again be important.'' Given limited knowledge at the time such an observation seems reasonable. The basic assumption that \pt\ spectra are monolithic is then confronted with apparently complex behavior such as the $x_t$  ``scaling'' trends noted above and detailed structure of exponent $\alpha(p_t)$. In contrast, introduction of a two-component model that is apparently {\em required} by spectrum data leads to major simplification such as factorization of charge-density and \pt\ dependences providing isolation of {\em fixed} power-law exponents as in Sec.~\ref{atrends}.

For the C-P energy regime global momentum conservation and the introduction of $x_t$ as a determining factor might have relevance. However, at RHIC and LHC energies behavior near midrapidity depends only indirectly on collision energy. For the 5 TeV data appearing in Fig.~\ref{direct} (d) above the maximum value of $x_t$ is approximately 0.007. The situation is perhaps clearer given the relation $\log(x_t) \sim y_t - y_z$. Spectrum structure near midrapidity at higher energies depends directly and simply on transverse rapidity \yt\ but only indirectly (parametrically) on longitudinal rapidity \yz\ (see Fig.~\ref{xxx} (a,c)). As a result of various attempts to arrive at spectrum ``scaling'' per various conjectured transformations the useful (i.e.\ interpretable) information is transformed into an uninterpretable parametrization. The C-P spectrum energy evolution is conveyed by Fig.~\ref{xxx} and Tables~\ref{2aanparamsq}, \ref{ppwidths} and \ref{pwwidthsxz}.

\subsection{Review of fixed-target experiments}

Reference~\cite{busza} is a review of fixed-target \pa\ collision data for beam energies at or below 400 GeV. One of the topics emphasized is high-\pt\ behavior of identified-hadron spectra vs hadron species and target A. The review is published in 1977 and so predates publication of Ref.~\cite{cronin0} that carries the bulk of information on \pa\ trends from the C-P collaboration. It is noted that \pa\ collisions may permit investigation of a projectile intermediate state passing through the target nucleus prior to formation of final-state hadrons, notable given discovery of ``nuclear transparency'' in the early seventies~\cite{busza3}.

Reference~\cite{busza} refers to Fig.~2 of Ref.~\cite{cronin10} (reproduced as its Fig.~15), a $\pi^-$ \pt\ spectrum for \pw\ collisions with 200, 300 and 400 GeV beam energies equivalent to Fig.~\ref{20bb} (c,d) of the present study. It notes that differential cross-section spectra scale $\propto A^{\alpha}$ and that $\alpha$ depends only on \pt\ and hadron species, not on collision energy (at least within the C-P energy range). In fact, Ref.~\cite{cronin10} (1974) emphasizes collision-energy dependence of \pw\ spectra and makes no reference to A dependence that is the main topic of Ref.~\cite{cronin0} (1979). Presumably this refers to preliminary C-P results made available for a review article.

Referring to its Fig.~17 Ref.~\cite{busza} questions why the $p/\pi^+$ ratio increases with A but {\em decreases} with $\sqrt{s}$. Response to that query illustrates the utility of the TCM for resolving such issues. Evidence for the first statement (Fig.~17a) reproduces Fig.~15 of Ref.~\cite{cronin0} (1979). The trend in that figure varies as $A^{a_{hp} - a_{h\pi^+}}$. Referring to Fig.~\ref{power} (d) above $a_{hp} \approx 0.6$ and $a_{h\pi^+} \approx 0.42$, so the corresponding trend in Fig.~15 of Ref.~\cite{cronin0} should be $\approx A^{0.18}$ compare to $\approx A^{0.16}$ derived from the data plotted in that figure,  within the  uncertainties of the present analysis.

Concerning variation of $p/\pi^+$ with $\sqrt{s}$, Fig.~17b of Ref.~\cite{busza} approximates Fig.~11 of Ref.~\cite{cronin10} which in turn corresponds to Fig.~\ref{n0x} (c) of  the present study. Note that while $p/\pi^+$ (upper points) decreases with $\sqrt{s}$, $\bar p/\pi^-$ (lower points) {\em increases}. In panel (c) above $y_t = 2$ ($p_t \approx 0.5$ GeV/c) $p/\pi^+$ data lie well above the soft-component ratio (dotted) and are thus controlled by hard-component width $\sigma_{y_t}$. In contrast, $\bar p/\pi^-$ data are below the soft-component ratio and are thus controlled by soft-component exponent $n$. Referring to Table~\ref{2aanparams} soft-component exponent $n$ decreases strongly with collision energy for $\bar p$ (compared to the $\pi^-$ trend) producing a strongly increasing $\bar p$ density and $\bar p/\pi^-$ ratio at higher \pt. In contrast, Table~\ref{pwwidthsxz} shows width variations for $p$ vs $\pi^+$ that are {\em relatively} stronger for the latter than the former leading to decrease with collision energy of  the {\em ratio} at higher \pt\ in  Fig.~17a of Ref.~\cite{busza}. Thus, while ratio $p/\pi^+$ is dominated by jet production ratio $\bar p/\pi^-$ is dominated by soft-component projectile fragmentation (and Gribov diffusion~\cite{gribov,gribov2}).

Note  that  species abundances at higher \pt\ depend on both integrated hard-component densities $\bar \rho_{hi}$ (common to all \pt\ intervals) and hard-component widths $\sigma_{y_ti}$ (dominating high-\pt\ densities). To good approximation \pt-{\em integrated} densities depend only on target A while TCM model parameters $n$ and $\sigma_{y_t}$ depend only on $\sqrt{s}$ (factorization). Spectrum ratios such as Fig.~17b of Ref.~\cite{busza} (Fig.~11 of Ref.~\cite{cronin10}) exaggerate the effect of model-parameter $\sqrt{s}$ variations on high-\pt\ yields misinterpreted to represent significant {\em integrated}-density A variations.

\section{Summary}\label{summ}

This study of fixed-target spectrum data from the Chicago-Princeton (C-P) collaboration obtained at the Fermi National Accelerator Laboratory (FNAL) relates to the importance of understanding small collision systems as references providing evidence for or against QGP formation in larger systems. In particular, target A dependence addressed by the C-P collaboration is of major interest as is the nature of the so-called Cronin effect.

C-P spectra are analyzed with a two-component (soft+hard) model of hadron production which facilitates distinction of jet (hard) and nonjet (soft) contributions and {\em factorization} of collision-energy $\sqrt{s}$ dependence and target-size A dependence for each component. Energy dependence for soft and hard \pt\ spectrum components is consistent with extrapolation of previous TCM results downward from LHC and RHIC energies (modulo the importance of valence quarks at fixed-target energies).

A major result of the present study is the finding that target A dependence for {\em individual} soft and hard particle densities consists of simple power laws $\bar \rho_s \propto A^{a_s}$ and $\bar \rho_h \propto A^{a_h}$ with {\em fixed exponents} $a_s$ and $a_h$ approximately independent of \pt\ and collision energy. \pt-dependent exponents $\alpha(p_t)$ reported by the C-P collaboration result from treating \pt\ spectra as monolithic, thus confusing factorizable \pt\ dependence and A dependence.

The soft-component exponent has fixed value $a_s \approx 0.20$ for several hadron species. Hard-component exponent $a_h \approx 0.40$ follows $a_h \approx 2a_s$ consistent with  expectations for individual \nn\ collisions for which $\bar \rho_h \propto \bar \rho_s^2$ is the rule and $N_{part}/2$ is close to 1 for {\em peripheral} minimum-bias or NSD \pa\ collisions. Interpreting A dependence then requires close examination of \pa\ centrality.

Conventional \pa\ centrality treatments based on the Glauber model greatly overestimate the number of nucleon participants $N_{part}$, binary collisions $N_{bin}$ and centrality (impact parameter) relative to measured particle production (e.g.\ charge multiplicity \nch). Systematic bias arises from two issues: large fluctuations of \nn\ particle production arising within projectile proton dissociation and {\em exclusivity} -- a given \nn\ interaction excludes other {\em simultaneous} interactions with either partner. \mbox{C-P} spectra correspond to minimum-bias (MB) event ensembles. A conventional Glauber analysis assigns a MB {\em average} as 50\% central. But an alternative analysis based on determining jet production (and hence binary collisions) via mean \mmpt\ data determines that an {\em individual} event (or narrow event class) consistent with the MB average is actually very peripheral, involving typically only one or two binary collisions (e.g. $N_{part}/2 \approx 1.2$ for \ppb).

Given those results data treatments that invoke $A^{1/3} \sim N_{bin}$ as an assumption may be quite misleading. Such methods typically involve spectrum ratios rescaled by $A^{1/3}$ (for particle density spectra) or $A$ (for differential cross sections). Rescaled ratios manifest A dependence in a hybrid form described as the ``Cronin effect'' wherein  {\em fixed} power-law exponents noted above are convoluted with undifferentiated hard and soft \pt\ spectrum components, the combination being uninterpretable on its own.



\begin{thebibliography}{99}

\bibitem{alippss} S.~Acharya \textit{et al.} (ALICE),
Eur. Phys. J. C \textbf{80}, no.2, 167 (2020).

\bibitem{alicestrange} J.~Adam \textit{et al.} (ALICE),
Nature Phys. \textbf{13}, 535-539 (2017).

\bibitem{cmsridge}   CMS~Collaboration,
  JHEP {\bf 1009}, 091 (2010).

\bibitem{phenix} C.~Aidala \textit{et al.} (PHENIX),
Nature Phys. \textbf{15}, no.3, 214-220 (2019)
doi:10.1038/s41567-018-0360-0
[arXiv:1805.02973 [nucl-ex]].

\bibitem{gardim} F.~G.~Gardim, R.~Krupczak and T.~N.~da Silva,
Phys. Rev. C \textbf{109}, no.1, 014904 (2024).

\bibitem{ppquad} T.~A.~Trainor and D.~J.~Prindle,
Phys. Rev. D \textbf{93}, no.1, 014031 (2016).

\bibitem{ppbpid}  T.~A.~Trainor,
J.\ Phys.\ G \textbf{47}, no.4, 045104 (2020).

\bibitem{pidpart1} T.~A.~Trainor,
arXiv:2112.09790.

\bibitem{pidpart2} T.~A.~Trainor,
arXiv:2112.12330.

\bibitem{ppsss} T.~A.~Trainor,
arXiv:2303.14299.

\bibitem{pppid} T.~A.~Trainor,
arXiv:2210.05877.

\bibitem{cronin10}  J.~W.~Cronin \textit{et al.} (E100),
Phys. Rev. D \textbf{11}, 3105-3123 (1975).


\bibitem{cronin0} D.~Antreasyan, J.~W.~Cronin, H.~J.~Frisch, M.~J.~Shochet, L.~Kluberg, P.~A.~Piroue and R.~L.~Sumner,
Phys. Rev. D \textbf{19}, 764-778 (1979).

\bibitem{croninabs} J.~W.~Cronin, H.~J.~Frisch, M.~J.~Shochet, J.~P.~Boymond, R.~Mermod, P.~A.~Piroue and R.~L.~Sumner,
PRINT-74-1182 (EFI,CHICAGO).

\bibitem{ppbnmf} T.~A.~Trainor,
arXiv:2304.02170.

\bibitem{ppprd} J.~Adams {\it et al.}  (STAR Collaboration),
Phys.\ Rev.\  D {\bf 74}, 032006 (2006).

\bibitem{newpptcm} T.~A.~Trainor,
arXiv:2104.08423.

\bibitem{wilk}  G.~Wilk and Z.~Wlodarczyk,
Phys.\ Rev.\ Lett.\  {\bf 84}, 2770 (2000).

\bibitem{gribov}  Y.~L.~Dokshitzer and D.~E.~Kharzeev,
Ann.\ Rev.\ Nucl.\ Part.\ Sci.\  {\bf 54}, 487 (2004).

\bibitem{gribov3} M.~L.~Nekrasov,
Particles \textbf{4}, no.3, 381-390 (2021).

\bibitem{gribov2} M.~L.~Nekrasov,
Mod. Phys. Lett. A \textbf{35}, no.38, 2050314 (2020).

\bibitem{hardspec}  T.~A.~Trainor,
Int.\ J.\ Mod.\ Phys.\  E {\bf 17}, 1499 (2008).

\bibitem{fragevo}    T.~A.~Trainor,
Phys.\ Rev.\  C {\bf 80}, 044901 (2009).

\bibitem{alicetomspec}    T.~A.~Trainor,
J.\ Phys.\ G {\bf 44}, no. 7, 075008 (2017).

\bibitem{alicenucmod} J.~Adam \textit{et al.} (ALICE),
Phys. Lett. B \textbf{760}, 720-735 (2016).

\bibitem{jetspec2}  T.~A.~Trainor,
Phys.\ Rev.\ D  {\bf 89}, 094011 (2014).

\bibitem{busza4} J.~E.~Elias, W.~Busza, C.~Halliwell, D.~Luckey, P.~Swartz, L.~Votta and C.~Young,
Phys. Rev. D \textbf{22}, 13 (1980).

\bibitem{tomglauber}  T.~A.~Trainor,
arXiv:1801.05862

\bibitem{aliceglauber} J.~Adam \textit{et al.} (ALICE),
Phys. Rev. C \textbf{91}, no.6, 064905 (2015).

\bibitem{tommpt} T.~A.~Trainor,
arXiv:1708.09412.

\bibitem{alicempt}  B.~B.~Abelev {\it et al.}  (ALICE Collaboration),
Phys.\ Lett.\ B {\bf 727}, 371 (2013).

\bibitem{aliceppmult}  J.~Adam {\it et al.} (ALICE Collaboration),
  Eur.\ Phys.\ J.\ C {\bf 77}, no. 1, 33 (2017).

\bibitem{nsdppp} V.~Khachatryan \textit{et al.} (CMS),
Phys. Rev. Lett. \textbf{105}, 022002 (2010).

\bibitem{alicensd} B.~Abelev \textit{et al.} (ALICE),
Phys. Rev. Lett. \textbf{110}, no.3, 032301 (2013).

\bibitem{tomexclude}  T.~A.~Trainor,
arXiv:1801.06579.

\bibitem{accardi} A.~Accardi,
hep-ph/0212148.

\bibitem{gunion} D.~L.~Jones and J.~F.~Gunion,
Phys. Rev. D \textbf{19}, 867 (1979).

\bibitem{busza} W.~Busza,
Acta Phys. Polon. B \textbf{8}, 333 (1977).

\bibitem{busza3} C.~Halliwell, J.~E.~Elias, W.~Busza, D.~Luckey, L.~Votta and C.~Young,
Phys. Rev. Lett. \textbf{39}, 1499-1502 (1977).


\end{thebibliography}
\end{document}